\documentclass[aps,prb,onecolumn,superscriptaddress,nofootinbib,floatfix,amsmath,amssymb]{revtex4-2}

\usepackage[utf8]{inputenc}
\usepackage{amsmath}
\usepackage{slashed}
\usepackage{graphicx}
\usepackage[caption=false]{subfig}
\usepackage{xcolor}
\usepackage[pdftex]{hyperref}
\begin{document}
\title{Effects of discrete topology on quantum transport across a graphene $n-p-n$ junction:\\ A quantum gravity analogue}
\author{Naveed Ahmad Shah}
		\email{naveed179755@st.jmi.ac.in}
		\affiliation{Department of Physics, Jamia Millia Islamia, New Delhi - 110025, India}
\author{Alonso Contreras-Astorga}
		\email{alonso.contreras@conacyt.mx}
		\affiliation{CONACyT-Physics Department, Cinvestav, P.O. Box. 14-740, 07000 Mexico City, Mexico}
		
\author{Fran\c{c}ois Fillion-Gourdeau}
		\email{francois.fillion@inrs.ca}
		\affiliation{Institute for Quantum Computing, University of Waterloo, Waterloo, Ontario, Canada, N2L 3G1}
		\affiliation{Infinite Potential Laboratories, Waterloo, Ontario, Canada, N2L 0A9}
\author{M. A. H Ahsan}
		\email{mahsan@jmi.ac.in}
		\affiliation{Department of Physics, Jamia Millia Islamia, New Delhi - 110025, India}
		
\author{Steve MacLean}
		\email{steve.maclean@inrs.ca}
		\affiliation{Institute for Quantum Computing, University of Waterloo, Waterloo, Ontario, Canada, N2L 3G1}
		\affiliation{Infinite Potential Laboratories, Waterloo, Ontario, Canada, N2L 0A9}
		\affiliation{Universit\'{e} du Qu\'{e}bec, INRS-\'{E}nergie, Mat\'{e}riaux et T\'{e}l\'{e}communications, Varennes, Qu\'{e}bec, Canada J3X 1S2}		
\author{Mir Faizal}
		\email{mirfaizalmir@googlemail.com}
		\affiliation{Irving K. Barber School of Arts and Sciences,
University of British Columbia - Okanagan, Kelowna, British Columbia V1V 1V7, Canada}
    \affiliation{Department of Physics and Astronomy,University of Lethbridge, Lethbridge, AB T1K 3M4, Canada}
	\affiliation{Canadian Quantum Research Center 204-3002  32 Ave Vernon, BC V1T 2L7 Canada}
\date{\today}

\begin{abstract}
In this article, we investigate the effect of next-to-the-nearest atom hopping on Klein tunnelling in graphene. An effective quantum dynamics equation is obtained based on an emergent generalized Dirac structure by analyzing the tight-binding model beyond the linear regime. We show that this structure has some interesting theoretical properties. First, it can be used to simplify quantum transport calculations used to characterize Klein tunnelling; second, it is not Chirally symmetric as hinted by previous work. Finally, it is reminiscent of theories on a space with a discrete topology. Exploiting these properties, we show that the discrete topology of the crystal lattice has an effect on the Klein tunnelling, which can be experimentally probed by measuring the transmittance through $n-p-n$ junctions. We argue that this simulates quantum gravitational analogues using graphene and we propose an experiment to perform such measurements.  
\end{abstract}

\maketitle

\section{Introduction}
At low energies, the quasi-particles of graphene responsible for its transport properties have a well-known description in terms of an emergent Dirac field theory \cite{PhysRevLett.53.2449,novoselov2005two,RevModPhys.81.109}. This property stems from the  symmetries of the underlying honeycomb 2D lattice, which reduces to a Dirac-like structure in the low-energy limit. The fact that the Dirac equation describes charge carriers makes graphene a testbed for relativistic theories in a nonrelativistic setting and it allows for the simulation of quantum electrodynamics (QED) in a condensed matter system \cite{KATSNELSON20073,PhysRevD.78.096009,PhysRevB.81.165431,PhysRevB.92.035401,PhysRevLett.124.110403,PhysRevB.78.085101,Wang734,PhysRevResearch.2.033472}.
Klein tunnelling was one of the first QED-like effects to be investigated in graphene \cite{PhysRevLett.102.026807,young2009quantum}. This relativistic-like phenomenon corresponds to the unimpeded transmission of particles through a potential step. Owing to the presence of negative energy states, there is no exponential damping and particles are almost fully transmitted when the potential barrier is larger than the rest mass energy of the particles. In graphene, conductance in transport experiments through $p-n$ junctions is the main observable used to detect this effect experimentally because Klein tunnelling has a strong effect on the transmittance  \cite{sonin2009effect,sanderson2013Klein,greenaway2015resonant,low2009conductance, perconte2018tunable}. The results of these experiments using electrostatic barriers were theoretically understood using the Dirac theory with the presence of negative energy states below the potential \cite{katsnelson2006chiral,allain2011Klein,RevModPhys.80.1337}. Although much work has been done in monolayer graphene \cite{avishai2021chiral, avishai2020Klein, da2019Klein, szabo2013relativistic, prada2010zero,Contreras-Astorga2020} and other condensed matter systems \cite{lee2019perfect}, the theoretical analysis were, for the most part, limited to the linear approximation (low energy limit) in the tight-binding model, where a standard Dirac structure exists. 

Some research on quantum transport has gone beyond this linear regime by studying various limits of the tight-binding model. In principle, these approaches are more accurate for the calculation of transport properties, especially at higher energies. For instance, implications of the next-to-nearest hopping term on doping and Klein tunnelling have been investigated in Ref.  \cite{PhysRevB.88.165427}. Within this approximation, it has been observed that the tunnelling is no longer Chiral and that an asymmetry occurs on conductance curves around the perfect transmission point. Furthermore, the effects of the trigonal warping terms \cite{PhysRevB.97.035420,Pereira_2010,PhysRevB.91.045420,  Pereira_Jr_2008} and deformed lattices \cite{PhysRevLett.104.063901} on transmittance has been considered. In particular, the Klein
tunnelling beyond the linear approximation has been studied using a generalized pseudospin mode-matching technique in the tight-binding model \cite{PhysRevB.97.035420}. This last approach shares many similarities with the technique presented in this paper.

In this article, we shall investigate Klein tunnelling beyond the linear approximation in monolayer graphene by including the next-to-the-nearest atom hopping. This is performed by introducing a generalized Dirac structure (GDS), which allows for  straightforward mode-matching calculations of the transmittance in the continuum limit. We argue that graphene beyond the linear regime can be used as a quantum gravity analogue. Indeed, the Dirac structure emerging in this limit is precisely of the kind obtained in quantum gravity models with a minimal length scenario \cite{doi:10.1142/S0218271818500803,amelino2013quantum,hossenfelder2013minimal,PhysRevLett.101.221301, PhysRevD.84.044013}. Based on this physical insight, we show that Klein tunnelling at intermediate energy is sensitive to the minimal length set by the lattice constant and thus, actually probes the underlying space topology. This effect simulates the effect of quantum gravity models on particle transport.  Finally, we evaluate the transmittance in a $n-p-n$ junction and we make a proposal to measure these effects experimentally. 
\section{Generalized Dirac structure from discrete topology of graphene}
\begin{figure} 
\includegraphics[width=0.3\textwidth]{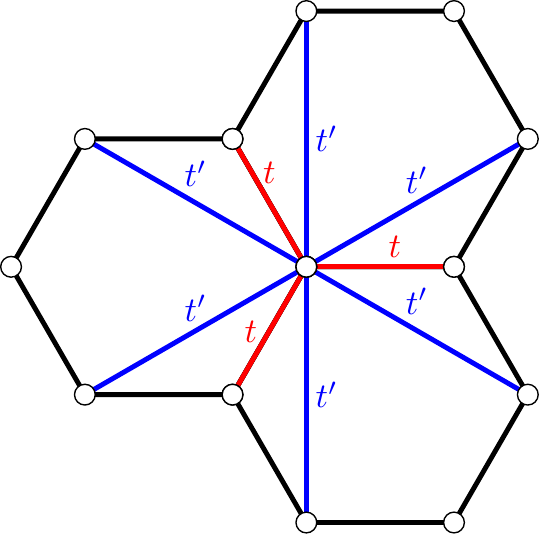}
\caption{Graphene hexagonal lattice with nearest ($t$) and next-to-nearest ($t'$) hopping energies.}
\label{fig:tbg}
\end{figure}

The effects of discrete topology of graphene can be investigated using its tight-binding model.
In a tight-binding model of graphene, the electron dispersion relation, expanded to $O((ka)^{2})$ is given by \cite{RevModPhys.81.109,RevModPhys.83.1193,PhysRevLett.100.087401, PhysRevB.88.165427}
\begin{eqnarray}
\label{eq:disp_full}
    \epsilon^{\lambda}_{\boldsymbol{k}} &=& 
    \lambda \hbar v_F k + \frac{9 a^2}{4} t^{'} k^2 
    - \lambda    \frac{\hbar v_F k^2 a }{4} \cos (3 \phi_{\boldsymbol{k}}), 
\end{eqnarray}
 where $\lambda = \pm 1$ is the band index, $v_F = 3ta/2\hbar$ is the Fermi velocity, $k = |\boldsymbol{k}|$ is the wavevector magnitude, $t,t'$ are the nearest and the next-to-nearest neighbour hopping energies (see Fig. \ref{fig:tbg}) while $a = 1.4 \; \mbox{\AA}$ is the lattice constant. Also, $\phi_{\boldsymbol{k}} = \arctan (k_{y}/k_{x})$ is the azymuthal angle of the wavevector with respect to the $x$-axis. The dispersion relation Eq. \eqref{eq:disp_full} comprises three different terms: the first term is the low energy linear contribution, the second one appears as the low energy limit of the next-to-nearest neighbour hopping and the third term, called trigonal warping, is the next-to-leading-order contribution obtained from the low energy limit of the tight-binding model with nearest-neighbour hopping. We can see from the above equation that the $t'$ correction breaks the electron-hole (Chiral) symmetry in the sense that $\epsilon^{-\lambda}_{\boldsymbol{k}} \neq - \epsilon^{\lambda}_{\boldsymbol{k}} $.
The dispersion relation for massless particles in many quantum gravity models is rotationally symmetric, of the form $\epsilon_{\mathrm{grav},\boldsymbol{p}} = F(p)$, where $F$ is usually a polynomial function and $\boldsymbol{p} = \hbar \boldsymbol{k}$ is now the momentum, related to the wave vector via De Broglie relations. One particular model has been extensively studied in which the dispersion relation is given by
$\epsilon_{\mathrm{grav},\boldsymbol{p}} = c_{1} p + c_{2} p^{2},$
where $c_{1,2}$ are some coefficients \cite{doi:10.1142/S0217751X97000566,Amelino2004,PhysRevD.62.053010}. This dispersion relation explicitly breaks Lorentz invariance even though it is rotationally invariant because it is not conserving the four-momentum of a particle.  

We make a connection between graphene and quantum gravity models by neglecting the trigonal warping term. Because this term encodes the symmetry of the lattice and breaks rotational invariance, it can be neglected for small angular regions around the angles $\phi^{\delta}_{\boldsymbol{k}} = (2n+1)\frac{\pi}{6} \pm \delta$ with $n \in \mathbb{N}$ and where $\delta= \frac{1}{3} \sin^{-1}(\frac{3|t'|}{5t}) \approx 0.02$. In this angular region, we have $\cos(3 \phi_{\boldsymbol{k}}) \approx 0$. Assuming all processes occur in a momentum region where we can neglect the trigonal warping term and renaming some variables, the energy becomes
\begin{eqnarray}
\epsilon^{\lambda}_{\boldsymbol{p}} =  v_F \left(\lambda p - \alpha p^2 \right), \label{dispersion}
\end{eqnarray}
where $\alpha = \frac{3}{2} \frac{|t^{'}|}{t} \frac{a}{\hbar} >0$. Obviously, this has the same form as $\epsilon_{\mathrm{grav},\boldsymbol{p}}$ with the connection provided by the mapping $c_{1} \rightarrow v_{F} \lambda$ and $c_{2} \rightarrow - v_{F} \alpha$.


To derive a dynamical Dirac-like equation having a dispersion relation given by Eq. \eqref{dispersion}, we borrow the technique developed for quantum gravity models with minimal lengths. As demonstrated in Ref. \cite{DAS2010407}, this can be performed by following the Dirac prescription $p \rightarrow \boldsymbol{\sigma} \cdot \hat{\boldsymbol{p}}$ in the expression for the energy, where $\hat{\boldsymbol{p}} = -i \hbar \nabla$ is now the momentum operator and $\boldsymbol{\sigma}$ are  Pauli matrices. We use the  Dirac prescription and  represent SO(2) invariant vectors in their spinor representation.
This yields the Hamiltonian operator given by:
\begin{eqnarray}
    \hat{H} &=& v_F \left[ \boldsymbol{\sigma} \cdot \hat{\boldsymbol{p}} - \alpha (\boldsymbol{\sigma} \cdot \hat{\boldsymbol{p}})  (\boldsymbol{\sigma} \cdot \hat{\boldsymbol{p}})\right] \nonumber \\
     &=& v_F \left[ \boldsymbol{\sigma} \cdot  \hat{\boldsymbol{p}} - \alpha \hat{p}^2 \sigma_0 \right] , \label{HAMIL} 
\end{eqnarray}
where $\sigma_{0} = \mathbb{I}_{2}$ is the unit matrix. This procedure reproduces the low energy limit of the next-to-nearest neighbour Hamiltonian contribution at next-to-leading order \cite{RevModPhys.83.1193}. To emphasise the fact that $\hat{H}$ has a GDS, the corresponding dynamical equation is now given:
\begin{eqnarray}
\label{eq:GDS}
i\partial_{t} \psi(t,\boldsymbol{x}) &=& v_F ( \boldsymbol{\sigma} \cdot  \tilde{\boldsymbol{P}} ) \psi(t,\boldsymbol{x}),
\end{eqnarray}
where $\tilde{\boldsymbol{P}} = \hat{\boldsymbol{p}} \sigma_{0} - \boldsymbol{\sigma} \alpha \hat{p}^2$ and where $\psi$ is the two-components spinor wavefunction. Equation \eqref{eq:GDS} is a generalized massless Dirac equation. In addition, it can be readily tested that the dispersion relation obtained from the energy eigenvalue equation for the above (Eq. \eqref{eveqn} in Appendix \ref{app:sol}) is the same as Eq. \eqref{dispersion}.

The mathematical structure defined thus far by the GDS in Eq.\eqref{eq:GDS} and the dispersion relation in Eq.\eqref{dispersion} is consistent with the framework of the generalized uncertainty principle (GUP). In this framework, one postulates the existence of generalized position and momentum operators $\hat{\boldsymbol{X}},\hat{\boldsymbol{P}}$ that obey a modified commutation relation (see Appendix \ref{app:gds}):
\begin{eqnarray}
\label{eq:comm}
[\hat{X}_{i},\hat{P}_{j}] = i\hbar \left[\delta_{ij} - \alpha \left(\delta_{ij} \hat{P} +\frac{\hat{P}_{i}\hat{P}_{j}}{\hat{P}} \right)\right].
\end{eqnarray}
These relations imply a minimal measurable length $(\Delta x)_{\mathrm{min}} \sim \hbar \alpha$ and a maximal measurable momentum $(\Delta p)_{\mathrm{min}} \sim \alpha^{-1}$ \cite{ALI2009497,DAS2010407}. Therefore, in this formalism, $\alpha \propto a$ is a parameter that captures the effect of the discrete topology. It appears in our description of graphene owing to the discreetness of its atomic structure.

To obey the generalized commutation relations \eqref{eq:comm}, the generalized position and momentum operators must be related to the usual position and momentum operators as 
$\hat{\boldsymbol{X}} = \hat{\boldsymbol{x}} $ and $ \hat{\boldsymbol{P}} = \hat{\boldsymbol{p}} (1 - \alpha \hat{p})$,
with $[\hat{x}_{i},\hat{p}_{j}] = i\hbar \delta_{ij}$. The operators $\hat{\boldsymbol{X}},\hat{\boldsymbol{P}}$ can be interpreted as the high-energy position and momentum, respectively, while $\hat{\boldsymbol{x}},\hat{\boldsymbol{p}}$ are their low-energy counterparts \cite{doi:10.1142/S0218271818500803}. The definition of the generalized momentum can be understood as a rotational-symmetric energy-scale transformation.  The Dirac-like equation obtained from this formalism has the form given in Eq. \eqref{eq:GDS} when one replaces the momentum operator by the generalized one and by using the Dirac prescription \cite{DAS2010407,doi:10.1142/S0218271818500803}.

As expected from the dispersion relation, the full Lorentz symmetry is not preserved in this mathematical structure and this implies particle/hole asymmetry. This symmetry breaking is algebraically indicated by the presence of $\hat{P}_{i}\hat{P}_{j}/\hat{P}$ terms in the commutation relations \eqref{eq:comm}. Thus, Chiral symmetry breaking arises from a more fundamental, although a less stringent, phenomenon of Lorentz symmetry breaking, which is reminiscent of the discrete topology of the space itself. It is less stringent, because we still consider the space to be isotropic (which shows up as the $SO(2)$ symmetry).

It is interesting to note that the algebraic structure induced by the GUP also occurs in some quantum gravity models \cite{doi:10.1142/S0218271818500803,amelino2013quantum,PhysRevLett.101.221301, PhysRevD.84.044013}, where the Planck length plays a role analogous to inter-atomic length in graphene. Actually, this formalism was developed to be consistent with string theory, black hole physics and doubly special
relativity \cite{ALI2009497,DAS2010407}. In gravitational theories, however, we have $\alpha \sim \ell_{\mathrm{plank}}/\hbar$, where $\ell_{\mathrm{plank}}$ is the Planck length. As a consequence, nonlinear corrections to the dispersion are very weak, unless one probes the system at Planck energy scales. This is not possible with actual experimental apparatus.   

On the other hand, nonlinear corrections in graphene start to be important at much lower energy scales \cite{PhysRevLett.100.087401}. Furthermore, as demonstrated in this article, it follows an algebraic structure consistent with GUP for some specific angles $\phi_{\boldsymbol{k}}$. Therefore, it is interesting to look at phenomenological implications of these corrections as they can be used as   
quantum gravity analogues. For this purpose, we analyze Klein tunnelling at intermediate energies ($\sim$ 100 meV) to investigate the effects of the $\alpha$-term on this phenomenon.

\section{Quantum transport with the generalized Dirac structure in graphene}
To study Klein tunnelling, we consider free waves scattering on a $n-p-n$ junction (more details of this calculation can be found in Appendix B). Thus, an electric static barrier potential $V(x)= V_0 \Theta(x)\Theta(D-x)$, where $D$ is the potential length, is introduced in the GDS resulting in   
\begin{align}
\hat{H} \  \Psi(\boldsymbol{r}) &= v_F \left[ - i \hbar \boldsymbol{\sigma} \cdot \nabla  + \alpha \hbar^2 \ \sigma_{0}\ \nabla^{2}+V(x)\right] \Psi(\boldsymbol{r})  = E_G \  \Psi(\boldsymbol{r}). \label{GDS equation}
\end{align}
Here (as in ordinary  Klein tunnelling \cite{katsnelson2006chiral,allain2011Klein}) the potential barrier distinguishes three different regions: \emph{region $A$} ($V(x) =   0, ~ x \leq 0$), with an incoming wave and a reflected one from the interface at $x=0$; \emph{region $B$} ($V(x) =   V_0,~  0 < x \leq D$), with two waves, one transmitted from region $A$ and another reflected by the interface at $x=D$; and \emph{region $C$} ($V(x) =  0$,  $x > D$), where there is just a transmitted wave. In the piecewise constant potential $V(x)$, Eq. \eqref{GDS equation} admits plane-wave
solutions $\Psi(\boldsymbol{x}) =  e^{i k_y y}\varphi(x)$, where:
%
%
\begin{align}
\varphi(x) = \begin{cases} 
e^{i k_x x} \begin{pmatrix} 1 \\ e^{i \phi} \end{pmatrix} + r  e^{-i k_x x} \begin{pmatrix} 1 \\ -e^{-i \phi} \end{pmatrix},    & x < 0; \\
a e^{i q_x x} \begin{pmatrix} 1 \\ -e^{i \theta} \end{pmatrix} + b   e^{-i q_x x} \begin{pmatrix} 1 \\ e^{-i \theta} \end{pmatrix}, & 0 < x \leq D ;\\
t  e^{i k_x x} \begin{pmatrix} 1 \\ e^{i \phi} \end{pmatrix}, & x > D ; \label{wavefs G}
\end{cases}
\end{align}
In the above, $r,t$ are the reflection and transmission coefficients, respectively, while $a,b$ are the coefficient of the waves under the barrier. 
Also, $\boldsymbol{k}$ is the wave vector in regions $A$ and $C$ while $\boldsymbol{q}$ is the wave vector in region $B$. The angles $\phi = \arctan (k_{y}/k_{x}), \theta = \arctan (q_{y}/q_{x})$ are the incident and transmitted angles, respectively. Since the potential  $V(x)$ is translationally invariant along the $y$-axis, the projection of the wavevector $k_y$ is a conserved quantity and we have $k_y = q_y$. Without loss of generality, we will assume $k_y>0$, representing an electron travelling from the lower half-plane $y<0$ to the upper half-plane $y>0$. Moreover, we will restrict ourselves to the energy range $0 < E_G < V_0$, where the Klein tunnelling phenomenon occurs. The energy $E_G$ in region $A$ is  related to $k$ via Eq. \eqref{dispersion}. Also, we can find $q_x$ in region $B$:
\begin{eqnarray} \label{Qx}
q_x^2 &=&  \frac{1}{\left(2 \alpha \hbar \right)^{2}} \Biggl( 2  - \frac{4 \alpha (E_G- V_0) }{v_F} - 2 \sqrt{1- \frac{4 \alpha (E_G- V_0) }{v_F}  } \Biggr)  - k^2\sin^2 \phi \label{q_x1}  .
\end{eqnarray}
%
The transmitted angle $\theta$ can be found using  $\tan \theta = \left( {k \sin \phi}/{q_x} \right)$.
Then matching the modes, i.e. equating the spinors in the three regions at the two boundaries of the potential assuming the continuity of the solution $\Psi(x,y)$ at $x=0$ and $x=D$, gives us the conditions on the coefficients of the spinors. 
Solving the resulting system of four simultaneous equations for $r$, and using $T_G = 1- |r|^2 $, we finally obtain the transmittance as 
\begin{eqnarray} \label{TG}
T_G &=& \frac{\cos^2 \phi \cos^2 \theta }{\big[\cos^2 (q_x D) \cos^2 \phi \cos^2 \theta + \sin^2 (q_x D) \left(1 + \sin \phi \sin \theta \right)^{2}\big]}.
\end{eqnarray}

For normal incidence, the transmission probability is $T_G=1$ and it is independent of the barrier height $V_0$ and  width $D$, like that in the linear case. The transmission probability becomes unity also when $q_x D= \pi n,$ $n \in \mathbb{Z}$.  

\begin{figure} 
\includegraphics[height=5.5 cm]{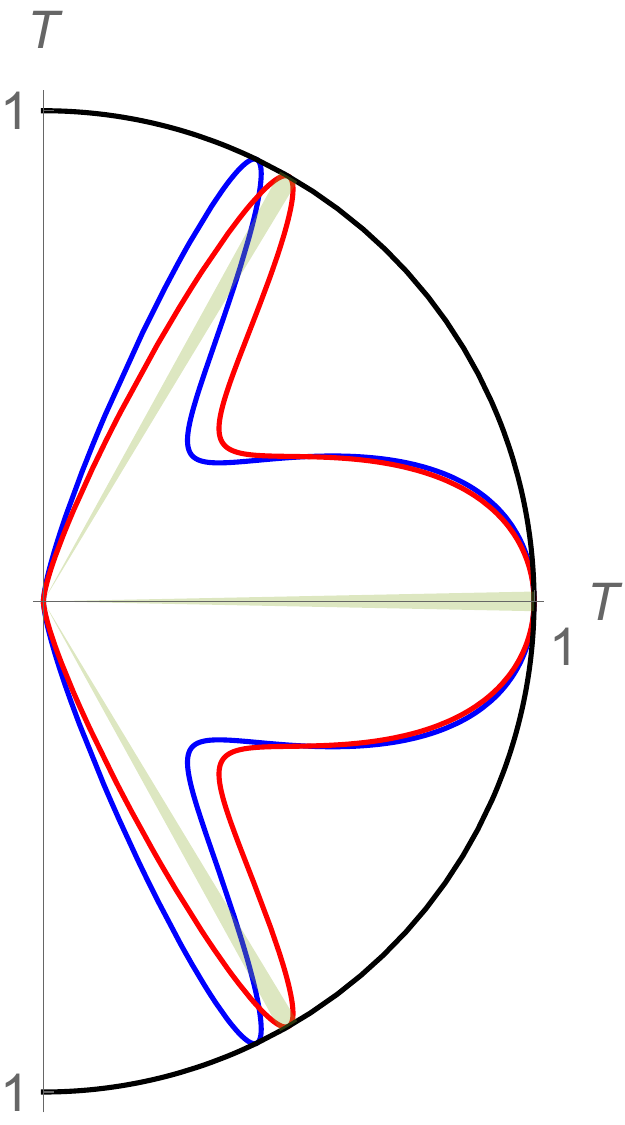} 
\hspace{0.5cm} \includegraphics[height=5.5 cm]{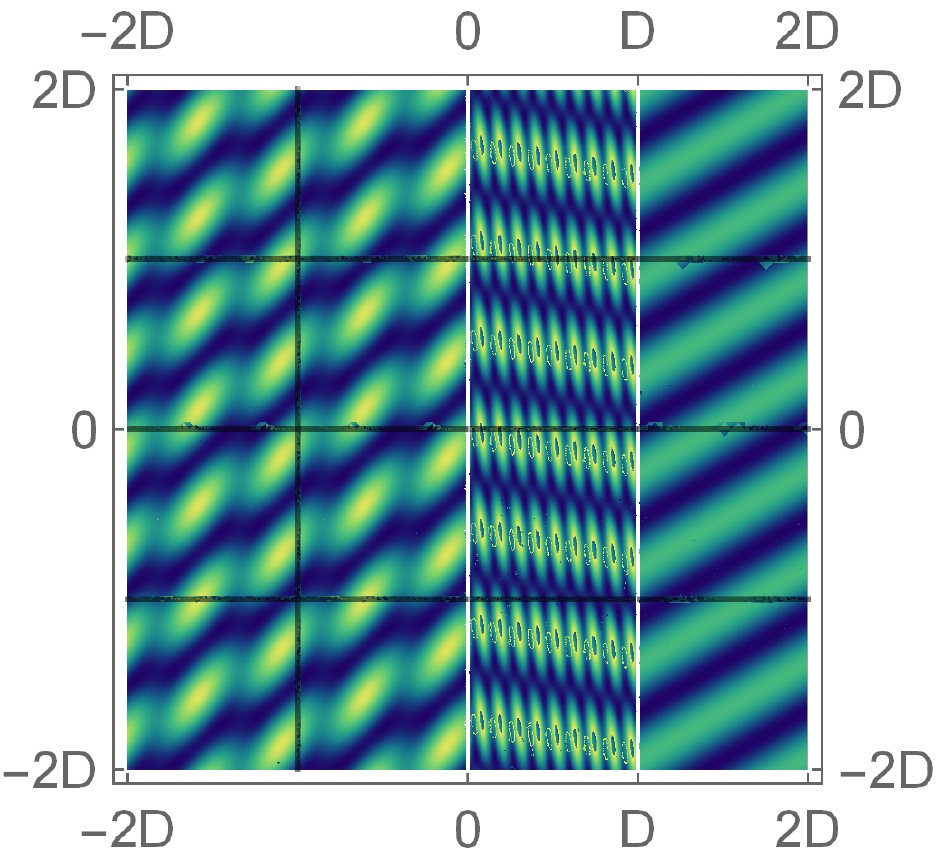}
\hspace{0.5cm} \includegraphics[height=5.5 cm]{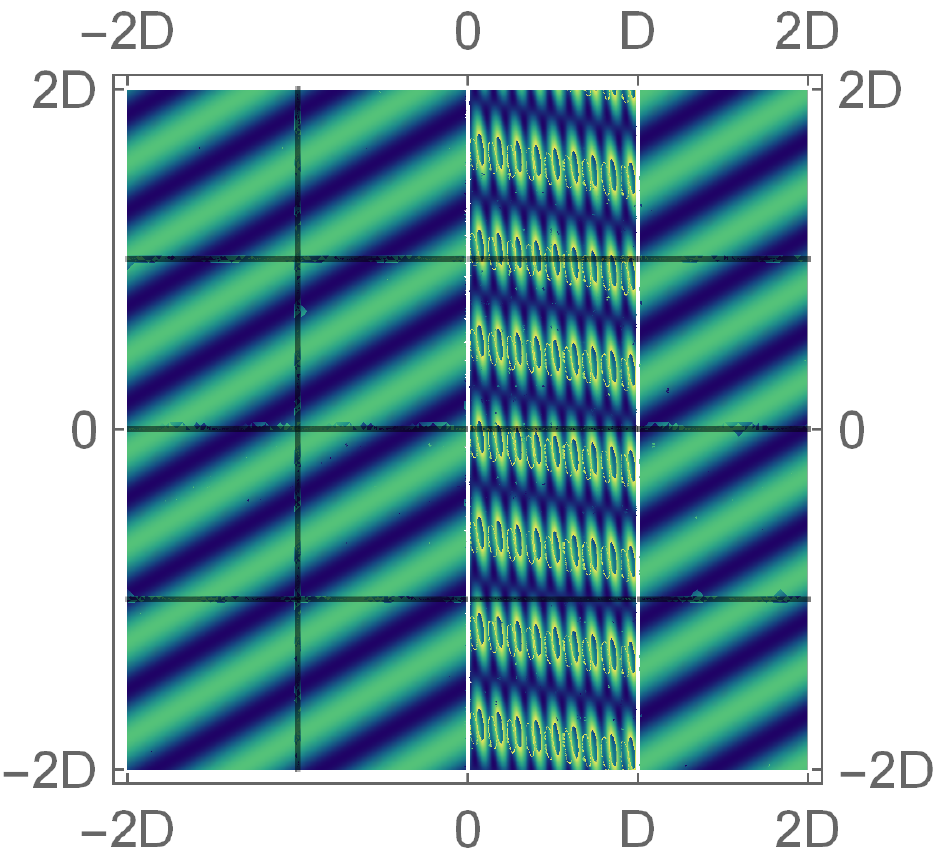}
\includegraphics[height=5.5 cm]{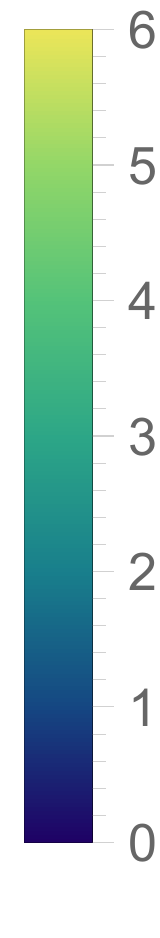}
\caption{
\textbf{Left}: Transmission probability $T$ and $T_G$ through a $V_0 = 285$ meV-high and $D=  96$ nm-wide  barrier as a function of the incidence angle $\phi$,  both with   $k=k_0=1.3407\times 10^8$ m$^{-1}$. The blue curve is the linear prediction of $T$ and the red curve the prediction $T_G$ with the GDS.  The black curve corresponds to unit transmission.  
\textbf{Center}: Probability density in the $x-y$ plane of the superposition of two plane waves, $\Phi= \Psi(k_0,0)+\Psi(k_0,\pi/3)$, considering only the linear regime. 
\textbf{Right}: Probability density $|| \Phi ||^2$ using the generalized Dirac structure. In all plots, the barrier is placed parallel to the armchair direction.}
\label{fig:interference}
\end{figure}

In Fig. \ref{fig:interference} Left, we show a polar plot of the transmission coefficient $T$ and $T_G$ to compare the predictions of the linear versus the generalized Dirac structures. The linear case is obtained by setting $\alpha = 0$.  The blue curve is the linear prediction of $T$ as in \cite{katsnelson2006chiral} and the red curve the prediction $T_G$ with the generalized Dirac structure  \eqref{TG}. The barrier is placed parallel to the armchair direction. The green-shadowed areas are regions for the incident angle $\phi$ where trigonal warping can be neglected.  In these areas, Eq. \eqref{GDS equation} represents correctly Klein tunelling up to $O((ka)^2)$. Since \eqref{GDS equation} is a linear differential equation, then it is also a correct representation in the subspace spanned by the plane waves with incident angle 
$\phi \in (-\delta,\delta) \cup (\pi/3-\delta, \pi/3 + \delta) \cup (-\pi/3 - \delta, -\pi/3+\delta)$.   Recall that we neglected the trigonal warping contribution; this is correct around $\phi^{\delta}_{\boldsymbol{k}}$ where we focus our analysis. There will be small corrections to Eq. \eqref{TG} at other angles. In Fig. \ref{fig:interference} Center and Right, we show the interference pattern produced by the superposition $\Phi= \Psi(k_0,0)+\Psi(k_0,\pi/3)$ considering the linear regime and the generalized Dirac structure, respectively. In the interference pattern in region $A$ (Center) there is a reflected wave while in the Right plot all the waves are transmitted from region $A$ through $C$.   
We considered the parameters of graphene reported in \cite{PhysRevB.88.165427}, $a=1.4$~\AA, $t=3 $ eV and $t'=-0.3$ eV. Moreover, the barrier is  $V_0 = 285$ meV high and $D=  96$ nm wide. The momentum  $k=k_0=1.3407\times 10^8$ m$^{-1}$ was selected using the condition $T_G=1$ at $\phi=0,\pm \pi/3$, where  trigonal warping is null. For $k_0$ the energy in the linear regime is $E=84.4638$ meV while considering the GDS is  $E_G = 84.226$ meV.

\section{Experimental proposal}
As demonstrated previously, it is possible to simulate the effect of minimum length on Klein tunnelling in graphene by showing that the transmittance $T_{G}(E,\phi)$, for some specific angles, is sensitive to the discrete topology ($T_{G} \neq T$). Probing this effect experimentally would be challenging because it requires a careful positioning of the potential barrier with respect to the lattice and because it requires resolving the transmission coefficient in angle and energy. However, this may be possible with actual experimental techniques. Indeed, time-resolved-photoemission electron microscopy (TR-PEEM) is capable of resolving the dynamics of charge carriers in time, space, energy and momentum with nanometer and femtosecond resolutions \cite{TUSCHE2015520}. In addition, laser beams can generate well-controlled photo-induced currents in graphene \cite{higuchi2017light}.
Therefore, we propose to use a pump-probe experiment where a mid-infrared pump photo-excites charge carriers and make them scatter on the potential barrier. Because this happens on short time scales (approximately the wave period of $\Delta t \sim 200$ fs), charge carriers travel large enough distances to go through the potential barrier ($\sim 200$ nm $\lesssim D$), but low enough to reduce possible scatterings with impurities and defects (ballistic regime). The laser field intensity required to accelerate charge carriers to the required energy is estimated as $I \sim 10^5$ W/cm$^2$, using $\Delta p \sim eE_{\ell}\Delta t$ where $E_{\ell}$ is the laser field strength. This is below the intensity of known mid-infrared laser sources \cite{Hafez_2016}. Furthermore, it is also possible to scan many scattering angles by modifying the laser polarization. The resulting dynamics can then be analyzed with the PEEM as a function of the pump-probe time-delay. Using the scattering dynamics of the distribution of charge carriers measured from the PEEM, we expect to be able to evaluate the transmission coefficient.


\section{Conclusion}
In this paper, we  have studied the effect of next-to-the-nearest atom hopping on quantum transport in graphene. This was   analytically done by  analyzing the tight-binding model beyond linear regime. The effective quantum dynamics obtained from such an approximation was used to    obtained  an emergent generalized Dirac structure, which captured the effects of discrete topology and which is reminiscent of Lorentz-breaking quantum gravity models. It was proposed that such effects can be tested by measuring transmittance through  $n-p-n$ graphene junctions using a pump-probe experiment. 

Here we will like to point out that the analogy with quantum gravity can be clearly seen by considering its analogy with  Doubly Special Relativity (DSR) \cite{amelino2013quantum}. In DSR, the Planck energy acts as the  maximal possible energy,  and   the smooth manifold structure of spacetime breaks beyond that energy. Furthermore, due to  DSR, the behavior of spacetime also changes at intermediate energies, and this can be  observed in effects like  the  breaking of  Lorentz symmetry.   In graphene, the situation is similar, as we have a maximum energy scale at which   the inter-atomic bounds break,  along with any smooth structure. This again have similar implications for intermediate  energy phenomena in graphene. Therefore, graphene can also be used as an analogue for Lorentz violating phenomena, which remain very elusive in high-energy particle physics experiments 
(see \cite{PhysRevLett.97.021601, doi:10.1142/S0219887821500481}).
It would be interesting to generalize this work by incorporating the effects of  a trigonal warping term. This could be done  by certain asymmetric expansion of the generalized momentum in terms of the standard low energy momentum. This will naturally furnish us a non-isotropic GUP reflecting the non-isotropy of space itself \cite{mann2021testing}.
It will be useful to use this formalism for the study of Klein tunnelling in graphene with a trigonal warping term. 

\begin{acknowledgments}
ACA acknowledges the support of Consejo Nacional de Ciencia  y  Tecnolog\'ia  (CONACyT-México)  under  the  grant FORDECYT-PRONACES/61533/2020. FFG and SM would like to thank Pierre L\'{e}vesque for discussions on the experimental proposal.
\end{acknowledgments}

\appendix
\section{Generalized Dirac structure in tight-binding model of graphene \label{app:gds}}

The energy dispersion relation in tight-binding model of graphene at intermediate energies beyond linear approximation is given by
\begin{equation}
\label{disp}
\epsilon^{\lambda}_{\boldsymbol{p}} =  v_F \left(\lambda p - \alpha p^2 \right).
\end{equation} 
Using the Dirac prescription $p \rightarrow \boldsymbol{\sigma} \cdot \hat{\boldsymbol{p}}$, we can express the Hamiltonian as 
  \begin{eqnarray}
      \hat{H} 
      &=& v_F[  \boldsymbol{\sigma} \cdot \hat{\boldsymbol{p}} - \alpha (\boldsymbol{\sigma} \cdot \hat{\boldsymbol{p}}) \ (\boldsymbol{\sigma} \cdot \hat{\boldsymbol{p}})] \nonumber \\
      &=& v_F \ \boldsymbol{\sigma} \cdot \hat{\boldsymbol{p}}(1 - \alpha \boldsymbol{\sigma} \cdot \hat{\boldsymbol{p}} )  
\end{eqnarray}
Adopting the definition of generalized momentum: $
\boldsymbol{\tilde{P}} = f(p)
$, where $f(p)$ is defined such that $\boldsymbol{\sigma} \cdot \boldsymbol{\tilde{P}} = \boldsymbol{\sigma} \cdot \hat{\boldsymbol{p}}(1 - \alpha \boldsymbol{\sigma} \cdot \hat{\boldsymbol{p}})$, 
the corresponding Hamiltonian can  be written in terms of the generalized momentum as 
\begin{eqnarray}
\hat{H} = v_F \ \boldsymbol{\sigma} \cdot \boldsymbol{\tilde{P}}. 
\end{eqnarray}
Let us consider now the operator  $ \hat{\boldsymbol{P}} = \hat{\boldsymbol{p}} (1 - \alpha \hat{p})$. Even though the generalized momentum $\hat{P}_i$ are functions of $\hat{p}_i$, we can choose $\hat{X}_i = \hat{x}_i$, where  $\hat{X}_{i}$ is the generalized coordinate conjugate  to $\hat{P}_{i}$. We assume  $\hat{x}_i$ and $\hat{p}_i$ to be the low-energy coordinates and momentum operators that satisfy the standard commutation relations $\left[\hat{x}_i , \hat{x}_j \right] = \left[\hat{p}_i , \hat{p}_j \right] = 0$, $\left[\hat{x}_i, \hat{p}_j\right] = i \hbar \delta_{ij}$. 
Using these low-energy coordinates and momentum commutator, we can write the  generalized commutation relations as 
\begin{eqnarray}
[\hat{X}_i, \hat{P}_j] 
&=& [\hat{x}_i, \hat{p}_j] - \alpha \left([\hat{x}_i, \hat{p}] \hat{p}_j + \hat{p} [\hat{x}_i, \hat{p}_j]\right)\nonumber \\
&=& i \hbar \delta_{ij} - i \alpha \hbar \left( \hat{p}_i \hat{p}^{-1} \ \hat{p}_j + \hat{p} \  \delta_{ij}\right).
 \end{eqnarray}
Now, we can express the low-energy momentum in terms of the generalized momentum, and obtain  
 \begin{eqnarray}
 [\hat{X}_i, \hat{P}_j]  &=& i \hbar \delta_{ij} - i \alpha \hbar \left[ (1 + \alpha \hat{P}) \hat{P}_i \ (1 + \alpha \hat{P})^{-1} \hat{P}^{-1} (1 + \alpha \hat{P}) \ \hat{P}_j \right] 
 - i \alpha \hbar \left[(1 + \alpha \hat{P}) \hat{P} \  \delta_{ij} \right].
 \end{eqnarray}
 Thus, neglecting $O(\alpha ^2)$ terms, we can write
 \begin{eqnarray}
 [\hat{X}_i, \hat{P}_j] 
 &=&  i \hbar \left[\delta_{ij} - \alpha \left(\hat{P} \delta_{ij} + \frac{\hat{P}_i \ \hat{P}_j}{\hat{P}}\right) \right]. \label{lsb}
\end{eqnarray}



\section{Solutions of the generalized Dirac equation \label{app:sol}}
Let us start from the generalized Dirac equation 
\begin{align}
\hat{H} \  \Psi(\boldsymbol{r}) = v_F \left[ - i \hbar \boldsymbol{\sigma} \cdot \nabla  + \alpha \hbar^2 \ \sigma_{0}\ \nabla^{2}+V(x)\right] \Psi(\boldsymbol{r}) = E_G \  \Psi(\boldsymbol{r}), \label{Sup: GDS equation}
\end{align}
where the potential barrier is the piecewise function  
\begin{eqnarray}
V(x) = \begin{cases} 
      0,        & 0 \leq x, \quad  \text{region } A; \\
      V_0,    & 0 \leq x \leq D, \text{region } B ;\\
      0, & x \geq D, \quad \text{region } C. 
       \end{cases}
\end{eqnarray}
Similar to the linear case, the solutions  are plane waves of the form  
\begin{equation} \label{Sup: ansatz}
\Psi(\boldsymbol{r}) = \exp({\pm i\boldsymbol{\kappa}\cdot \boldsymbol{r}})  \left(  \begin{array}{l}
 \xi_1 \\
\xi_2\\
\end{array} \right), 
\end{equation}
where $\boldsymbol{\kappa}$ is the wavevector in each region, $\boldsymbol{r} = (x,y)$ are the Cartesian coordinates, and $\xi_1 , \xi_2 $ are the components of the Dirac spinor. We first separate variables using the translational-invariance symmetry, $\Psi(\boldsymbol{x}) =  e^{i k_y y}\varphi(x)$. The components $\xi_1 , \xi_2 $ in each region are found by direct substitution of the ansatz in \eqref{Sup: GDS equation}.   Then, the ansatz \eqref{Sup: ansatz} can be written as 
\begin{eqnarray} \label{Sup: varphi}
\varphi(x) = \begin{cases} 
          e^{i k_x x} \begin{pmatrix} 1 \\ e^{i \phi} \end{pmatrix} + r \ e^{-i k_x x} \begin{pmatrix} 1 \\ -e^{-i \phi} \end{pmatrix},    & x \leq 0; \\
          a \ e^{i q_x x} \begin{pmatrix} 1 \\ -e^{i \theta} \end{pmatrix} + b \  e^{-i q_x x} \begin{pmatrix} 1 \\ e^{-i \theta} \end{pmatrix}, & 0 < x \leq D ;\\
          t \ e^{i k_x x} \begin{pmatrix} 1 \\ e^{i \phi} \end{pmatrix}, & x > D \label{wavefsGG}
       \end{cases}
\end{eqnarray}
where we expressed the wavevector $\boldsymbol{\kappa}$ in each region with the variables $\boldsymbol{\kappa} \rightarrow \boldsymbol{k}=(k_x, k_y)$, in regions $A$ and $C$; and $\boldsymbol{\kappa} \rightarrow \boldsymbol{q}=(q_x,q_y)$ in the region $0<x<D$.  In the above equation, $\phi = \tan^{-1}(k_y/k_x)$ is the incident angle of a wave traveling from left to right in the $x$-direction and $\theta = \tan^{-1}(q_y/q_x) = \tan^{-1}(k_y/q_x)$ is the angle of refraction. 

Now, from the eigenvalue equation $\hat{H} \Psi = E_G \Psi$ in region $x<0$, we obtain the relation between $E_G$ and $k$: 
\begin{eqnarray} 
E_G \Psi  &=& \hat{H} \Psi \nonumber \\
&=& -i v_F\hbar  \sigma_1 \partial_x \Psi -iv_F \hbar \sigma_2 \partial_y \Psi + \alpha v_F \hbar^2 \sigma_0 \left(\partial_x^2 + \partial_y^2  \right) \Psi  \nonumber \\
&=& \hbar v_F k \Psi - \alpha v_F \hbar^2 k^2 \Psi,
\end{eqnarray}
thus 
\begin{equation}
E_G =\hbar v_F k - \alpha v_F \hbar^2 k^2,  \label{eveqn}
\end{equation}
i.e., for a given energy $E_G$, the magnitude of the wavevector reads
\begin{eqnarray}
k = \frac{1}{2\alpha \hbar} \left(1 \pm \sqrt{1 - \frac{4\alpha E_G}{v_F}}\right). 
\end{eqnarray} 
Then, $\boldsymbol{k}=k(\cos \phi, \sin \phi)$. The same result applies to region $C$. 

Moreover, we can repeat the calculations for the region $0<x<D$ to find $\boldsymbol{q}$. Recall that there is conservation of momentum in the $y$-direction, then $q_y = k_y$. From  the Dirac equation,  we obtain: 
\begin{eqnarray}
E_G  \Psi &=& \hat{H} \Psi \nonumber \\ 
&=& -i v_F\hbar  \sigma_1 \partial_x \Psi -iv_F \hbar \sigma_2 \partial_y \Psi + \alpha v_F \hbar^2 \sigma_0 \left(\partial_x^2 + \partial_y^2  \right) \Psi + V_0 \Psi  \nonumber \\ 
&=& - \hbar v_F q \Psi - \alpha v_F \hbar^2 q^2 \Psi+V_0 \Psi,  
\end{eqnarray}
therfore, $E_G= -\hbar v_F q - \alpha v_F \hbar^2 q^2 +V_0$. To find $q_x$, we first observe that $q$ is given by 
\begin{eqnarray}
q = \frac{1}{2 \alpha \hbar}\left(-1 \pm \nu  \right), 
\end{eqnarray}
where $\nu^2 = 1 - {4 \alpha (E_G- V_0) }{v_F^{-1}}$. Note that $2 \alpha \hbar q +1 = \pm \nu  $, however   $ 2 \alpha \hbar q > 0$ and  $\nu > 0$, hence the physical solution is  $ q = {(2 \alpha \hbar)^{-1}}(-1 + \nu )$. We can obtain  $q_x$, using $q_x^2  = q^2-q_y^2 = q^2 - k_y^2$  
\begin{eqnarray}
q^2_x &=&  \left(\frac{1}{2 \alpha \hbar}\right)^2 \left( 2 - \frac{4 \alpha (E_G- V_0) }{v_F} - 2 \sqrt{1 - \frac{4 \alpha (E_G- V_0) }{v_F} } \right) \nonumber - k^2\sin^2 \phi.  \label{q_x11}
\end{eqnarray}

Finally, we ask the wavefunction $\Psi(\boldsymbol{r})$ to be continuous at $x=0$ and $x=D$. These conditions give a system of four linear equations that allow us to calculate the coefficients $a,~b,~r,~t$ of \eqref{Sup: varphi}. 
\bibliography{References}

\begin{thebibliography}{52}%
\makeatletter
\providecommand \@ifxundefined [1]{%
 \@ifx{#1\undefined}
}%
\providecommand \@ifnum [1]{%
 \ifnum #1\expandafter \@firstoftwo
 \else \expandafter \@secondoftwo
 \fi
}%
\providecommand \@ifx [1]{%
 \ifx #1\expandafter \@firstoftwo
 \else \expandafter \@secondoftwo
 \fi
}%
\providecommand \natexlab [1]{#1}%
\providecommand \enquote  [1]{``#1''}%
\providecommand \bibnamefont  [1]{#1}%
\providecommand \bibfnamefont [1]{#1}%
\providecommand \citenamefont [1]{#1}%
\providecommand \href@noop [0]{\@secondoftwo}%
\providecommand \href [0]{\begingroup \@sanitize@url \@href}%
\providecommand \@href[1]{\@@startlink{#1}\@@href}%
\providecommand \@@href[1]{\endgroup#1\@@endlink}%
\providecommand \@sanitize@url [0]{\catcode `\\12\catcode `\$12\catcode
  `\&12\catcode `\#12\catcode `\^12\catcode `\_12\catcode `\%12\relax}%
\providecommand \@@startlink[1]{}%
\providecommand \@@endlink[0]{}%
\providecommand \url  [0]{\begingroup\@sanitize@url \@url }%
\providecommand \@url [1]{\endgroup\@href {#1}{\urlprefix }}%
\providecommand \urlprefix  [0]{URL }%
\providecommand \Eprint [0]{\href }%
\providecommand \doibase [0]{https://doi.org/}%
\providecommand \selectlanguage [0]{\@gobble}%
\providecommand \bibinfo  [0]{\@secondoftwo}%
\providecommand \bibfield  [0]{\@secondoftwo}%
\providecommand \translation [1]{[#1]}%
\providecommand \BibitemOpen [0]{}%
\providecommand \bibitemStop [0]{}%
\providecommand \bibitemNoStop [0]{.\EOS\space}%
\providecommand \EOS [0]{\spacefactor3000\relax}%
\providecommand \BibitemShut  [1]{\csname bibitem#1\endcsname}%
\let\auto@bib@innerbib\@empty
\bibitem [{\citenamefont {Semenoff}(1984)}]{PhysRevLett.53.2449}%
  \BibitemOpen
  \bibfield  {author} {\bibinfo {author} {\bibfnamefont {G.~W.}\ \bibnamefont
  {Semenoff}},\ }\bibfield  {title} {\bibinfo {title} {Condensed-matter
  simulation of a three-dimensional anomaly},\ }\href
  {https://doi.org/10.1103/PhysRevLett.53.2449} {\bibfield  {journal} {\bibinfo
   {journal} {Phys. Rev. Lett.}\ }\textbf {\bibinfo {volume} {53}},\ \bibinfo
  {pages} {2449} (\bibinfo {year} {1984})}\BibitemShut {NoStop}%
\bibitem [{\citenamefont {Novoselov}\ \emph {et~al.}(2005)\citenamefont
  {Novoselov}, \citenamefont {Geim}, \citenamefont {Morozov}, \citenamefont
  {Jiang}, \citenamefont {Katsnelson}, \citenamefont {Grigorieva},
  \citenamefont {Dubonos},\ and\ \citenamefont {A}}]{novoselov2005two}%
  \BibitemOpen
  \bibfield  {author} {\bibinfo {author} {\bibfnamefont {K.~S.}\ \bibnamefont
  {Novoselov}}, \bibinfo {author} {\bibfnamefont {A.~K.}\ \bibnamefont {Geim}},
  \bibinfo {author} {\bibfnamefont {S.~V.}\ \bibnamefont {Morozov}}, \bibinfo
  {author} {\bibfnamefont {D.}~\bibnamefont {Jiang}}, \bibinfo {author}
  {\bibfnamefont {M.~I.}\ \bibnamefont {Katsnelson}}, \bibinfo {author}
  {\bibfnamefont {I.}~\bibnamefont {Grigorieva}}, \bibinfo {author}
  {\bibfnamefont {F.}~\bibnamefont {Dubonos}, \bibfnamefont {SVb}},\ and\
  \bibinfo {author} {\bibfnamefont {A.}~\bibnamefont {A}},\ }\bibfield  {title}
  {\bibinfo {title} {Two-dimensional gas of massless {D}irac fermions in
  graphene},\ }\href {https://doi.org/https://doi.org/10.1038/nature04233}
  {\bibfield  {journal} {\bibinfo  {journal} {Nature}\ }\textbf {\bibinfo
  {volume} {438}},\ \bibinfo {pages} {197} (\bibinfo {year}
  {2005})}\BibitemShut {NoStop}%
\bibitem [{\citenamefont {Castro~Neto}\ \emph {et~al.}(2009)\citenamefont
  {Castro~Neto}, \citenamefont {Guinea}, \citenamefont {Peres}, \citenamefont
  {Novoselov},\ and\ \citenamefont {Geim}}]{RevModPhys.81.109}%
  \BibitemOpen
  \bibfield  {author} {\bibinfo {author} {\bibfnamefont {A.~H.}\ \bibnamefont
  {Castro~Neto}}, \bibinfo {author} {\bibfnamefont {F.}~\bibnamefont {Guinea}},
  \bibinfo {author} {\bibfnamefont {N.~M.~R.}\ \bibnamefont {Peres}}, \bibinfo
  {author} {\bibfnamefont {K.~S.}\ \bibnamefont {Novoselov}},\ and\ \bibinfo
  {author} {\bibfnamefont {A.~K.}\ \bibnamefont {Geim}},\ }\bibfield  {title}
  {\bibinfo {title} {The electronic properties of graphene},\ }\href
  {https://doi.org/10.1103/RevModPhys.81.109} {\bibfield  {journal} {\bibinfo
  {journal} {Rev. Mod. Phys.}\ }\textbf {\bibinfo {volume} {81}},\ \bibinfo
  {pages} {109} (\bibinfo {year} {2009})}\BibitemShut {NoStop}%
\bibitem [{\citenamefont {Katsnelson}\ and\ \citenamefont
  {Novoselov}(2007)}]{KATSNELSON20073}%
  \BibitemOpen
  \bibfield  {author} {\bibinfo {author} {\bibfnamefont {M.~I.}\ \bibnamefont
  {Katsnelson}}\ and\ \bibinfo {author} {\bibfnamefont {K.~S.}\ \bibnamefont
  {Novoselov}},\ }\bibfield  {title} {\bibinfo {title} {Graphene: New bridge
  between condensed matter physics and quantum electrodynamics},\ }\href
  {https://doi.org/https://doi.org/10.1016/j.ssc.2007.02.043} {\bibfield
  {journal} {\bibinfo  {journal} {Solid State Commun.}\ }\textbf {\bibinfo
  {volume} {143}},\ \bibinfo {pages} {3} (\bibinfo {year} {2007})},\ \bibinfo
  {note} {exploring graphene}\BibitemShut {NoStop}%
\bibitem [{\citenamefont {Allor}\ \emph {et~al.}(2008)\citenamefont {Allor},
  \citenamefont {Cohen},\ and\ \citenamefont {McGady}}]{PhysRevD.78.096009}%
  \BibitemOpen
  \bibfield  {author} {\bibinfo {author} {\bibfnamefont {D.}~\bibnamefont
  {Allor}}, \bibinfo {author} {\bibfnamefont {T.~D.}\ \bibnamefont {Cohen}},\
  and\ \bibinfo {author} {\bibfnamefont {D.~A.}\ \bibnamefont {McGady}},\
  }\bibfield  {title} {\bibinfo {title} {Schwinger mechanism and graphene},\
  }\href {https://doi.org/10.1103/PhysRevD.78.096009} {\bibfield  {journal}
  {\bibinfo  {journal} {Phys. Rev. D}\ }\textbf {\bibinfo {volume} {78}},\
  \bibinfo {pages} {096009} (\bibinfo {year} {2008})}\BibitemShut {NoStop}%
\bibitem [{\citenamefont {D\'ora}\ and\ \citenamefont
  {Moessner}(2010)}]{PhysRevB.81.165431}%
  \BibitemOpen
  \bibfield  {author} {\bibinfo {author} {\bibfnamefont {B.}~\bibnamefont
  {D\'ora}}\ and\ \bibinfo {author} {\bibfnamefont {R.}~\bibnamefont
  {Moessner}},\ }\bibfield  {title} {\bibinfo {title} {Nonlinear electric
  transport in graphene: Quantum quench dynamics and the {S}chwinger
  mechanism},\ }\href {https://doi.org/10.1103/PhysRevB.81.165431} {\bibfield
  {journal} {\bibinfo  {journal} {Phys. Rev. B}\ }\textbf {\bibinfo {volume}
  {81}},\ \bibinfo {pages} {165431} (\bibinfo {year} {2010})}\BibitemShut
  {NoStop}%
\bibitem [{\citenamefont {Fillion-Gourdeau}\ and\ \citenamefont
  {MacLean}(2015)}]{PhysRevB.92.035401}%
  \BibitemOpen
  \bibfield  {author} {\bibinfo {author} {\bibfnamefont {F.}~\bibnamefont
  {Fillion-Gourdeau}}\ and\ \bibinfo {author} {\bibfnamefont {S.}~\bibnamefont
  {MacLean}},\ }\bibfield  {title} {\bibinfo {title} {Time-dependent pair
  creation and the {S}chwinger mechanism in graphene},\ }\href
  {https://doi.org/10.1103/PhysRevB.92.035401} {\bibfield  {journal} {\bibinfo
  {journal} {Phys. Rev. B}\ }\textbf {\bibinfo {volume} {92}},\ \bibinfo
  {pages} {035401} (\bibinfo {year} {2015})}\BibitemShut {NoStop}%
\bibitem [{\citenamefont {Golub}\ \emph {et~al.}(2020)\citenamefont {Golub},
  \citenamefont {Egger}, \citenamefont {M\"uller},\ and\ \citenamefont
  {Villalba-Ch\'avez}}]{PhysRevLett.124.110403}%
  \BibitemOpen
  \bibfield  {author} {\bibinfo {author} {\bibfnamefont {A.}~\bibnamefont
  {Golub}}, \bibinfo {author} {\bibfnamefont {R.}~\bibnamefont {Egger}},
  \bibinfo {author} {\bibfnamefont {C.}~\bibnamefont {M\"uller}},\ and\
  \bibinfo {author} {\bibfnamefont {S.}~\bibnamefont {Villalba-Ch\'avez}},\
  }\bibfield  {title} {\bibinfo {title} {Dimensionality-driven photoproduction
  of massive {D}irac pairs near threshold in gapped graphene monolayers},\
  }\href {https://doi.org/10.1103/PhysRevLett.124.110403} {\bibfield  {journal}
  {\bibinfo  {journal} {Phys. Rev. Lett.}\ }\textbf {\bibinfo {volume} {124}},\
  \bibinfo {pages} {110403} (\bibinfo {year} {2020})}\BibitemShut {NoStop}%
\bibitem [{\citenamefont {Pereira}\ \emph {et~al.}(2008)\citenamefont
  {Pereira}, \citenamefont {Kotov},\ and\ \citenamefont
  {Castro~Neto}}]{PhysRevB.78.085101}%
  \BibitemOpen
  \bibfield  {author} {\bibinfo {author} {\bibfnamefont {V.~M.}\ \bibnamefont
  {Pereira}}, \bibinfo {author} {\bibfnamefont {V.~N.}\ \bibnamefont {Kotov}},\
  and\ \bibinfo {author} {\bibfnamefont {A.~H.}\ \bibnamefont {Castro~Neto}},\
  }\bibfield  {title} {\bibinfo {title} {Supercritical {C}oulomb impurities in
  gapped graphene},\ }\href {https://doi.org/10.1103/PhysRevB.78.085101}
  {\bibfield  {journal} {\bibinfo  {journal} {Phys. Rev. B}\ }\textbf {\bibinfo
  {volume} {78}},\ \bibinfo {pages} {085101} (\bibinfo {year}
  {2008})}\BibitemShut {NoStop}%
\bibitem [{\citenamefont {Wang}\ \emph {et~al.}(2013)\citenamefont {Wang},
  \citenamefont {Wong}, \citenamefont {Shytov}, \citenamefont {Brar},
  \citenamefont {Choi}, \citenamefont {Wu}, \citenamefont {Tsai}, \citenamefont
  {Regan}, \citenamefont {Zettl}, \citenamefont {Kawakami}, \citenamefont
  {Louie}, \citenamefont {Levitov},\ and\ \citenamefont {Crommie}}]{Wang734}%
  \BibitemOpen
  \bibfield  {author} {\bibinfo {author} {\bibfnamefont {Y.}~\bibnamefont
  {Wang}}, \bibinfo {author} {\bibfnamefont {D.}~\bibnamefont {Wong}}, \bibinfo
  {author} {\bibfnamefont {A.~V.}\ \bibnamefont {Shytov}}, \bibinfo {author}
  {\bibfnamefont {V.~W.}\ \bibnamefont {Brar}}, \bibinfo {author}
  {\bibfnamefont {S.}~\bibnamefont {Choi}}, \bibinfo {author} {\bibfnamefont
  {Q.}~\bibnamefont {Wu}}, \bibinfo {author} {\bibfnamefont {H.-Z.}\
  \bibnamefont {Tsai}}, \bibinfo {author} {\bibfnamefont {W.}~\bibnamefont
  {Regan}}, \bibinfo {author} {\bibfnamefont {A.}~\bibnamefont {Zettl}},
  \bibinfo {author} {\bibfnamefont {R.~K.}\ \bibnamefont {Kawakami}}, \bibinfo
  {author} {\bibfnamefont {S.~G.}\ \bibnamefont {Louie}}, \bibinfo {author}
  {\bibfnamefont {L.~S.}\ \bibnamefont {Levitov}},\ and\ \bibinfo {author}
  {\bibfnamefont {M.~F.}\ \bibnamefont {Crommie}},\ }\bibfield  {title}
  {\bibinfo {title} {Observing atomic collapse resonances in artificial nuclei
  on graphene},\ }\href {https://doi.org/10.1126/science.1234320} {\bibfield
  {journal} {\bibinfo  {journal} {Science}\ }\textbf {\bibinfo {volume}
  {340}},\ \bibinfo {pages} {734} (\bibinfo {year} {2013})}\BibitemShut
  {NoStop}%
\bibitem [{\citenamefont {Fillion-Gourdeau}\ \emph {et~al.}(2020)\citenamefont
  {Fillion-Gourdeau}, \citenamefont {Levesque},\ and\ \citenamefont
  {MacLean}}]{PhysRevResearch.2.033472}%
  \BibitemOpen
  \bibfield  {author} {\bibinfo {author} {\bibfnamefont {F.}~\bibnamefont
  {Fillion-Gourdeau}}, \bibinfo {author} {\bibfnamefont {P.}~\bibnamefont
  {Levesque}},\ and\ \bibinfo {author} {\bibfnamefont {S.}~\bibnamefont
  {MacLean}},\ }\bibfield  {title} {\bibinfo {title} {Plunging in the dirac sea
  using graphene quantum dots},\ }\href
  {https://doi.org/10.1103/PhysRevResearch.2.033472} {\bibfield  {journal}
  {\bibinfo  {journal} {Phys. Rev. Research}\ }\textbf {\bibinfo {volume}
  {2}},\ \bibinfo {pages} {033472} (\bibinfo {year} {2020})}\BibitemShut
  {NoStop}%
\bibitem [{\citenamefont {Stander}\ \emph {et~al.}(2009)\citenamefont
  {Stander}, \citenamefont {Huard},\ and\ \citenamefont
  {Goldhaber-Gordon}}]{PhysRevLett.102.026807}%
  \BibitemOpen
  \bibfield  {author} {\bibinfo {author} {\bibfnamefont {N.}~\bibnamefont
  {Stander}}, \bibinfo {author} {\bibfnamefont {B.}~\bibnamefont {Huard}},\
  and\ \bibinfo {author} {\bibfnamefont {D.}~\bibnamefont {Goldhaber-Gordon}},\
  }\bibfield  {title} {\bibinfo {title} {Evidence for {K}lein tunneling in
  graphene $p\mathrm{\text{\ensuremath{-}}}n$ junctions},\ }\href
  {https://doi.org/10.1103/PhysRevLett.102.026807} {\bibfield  {journal}
  {\bibinfo  {journal} {Phys. Rev. Lett.}\ }\textbf {\bibinfo {volume} {102}},\
  \bibinfo {pages} {026807} (\bibinfo {year} {2009})}\BibitemShut {NoStop}%
\bibitem [{\citenamefont {Young}\ and\ \citenamefont
  {Kim}(2009)}]{young2009quantum}%
  \BibitemOpen
  \bibfield  {author} {\bibinfo {author} {\bibfnamefont {A.~F.}\ \bibnamefont
  {Young}}\ and\ \bibinfo {author} {\bibfnamefont {P.}~\bibnamefont {Kim}},\
  }\bibfield  {title} {\bibinfo {title} {Quantum interference and {K}lein
  tunnelling in graphene heterojunctions},\ }\href
  {https://doi.org/https://doi.org/10.1038/nphys1198} {\bibfield  {journal}
  {\bibinfo  {journal} {Nat. Phys.}\ }\textbf {\bibinfo {volume} {5}},\
  \bibinfo {pages} {222} (\bibinfo {year} {2009})}\BibitemShut {NoStop}%
\bibitem [{\citenamefont {Sonin}(2009)}]{sonin2009effect}%
  \BibitemOpen
  \bibfield  {author} {\bibinfo {author} {\bibfnamefont {E.}~\bibnamefont
  {Sonin}},\ }\bibfield  {title} {\bibinfo {title} {Effect of {K}lein tunneling
  on conductance and shot noise in ballistic graphene},\ }\href
  {https://doi.org/10.1103/PhysRevB.79.195438} {\bibfield  {journal} {\bibinfo
  {journal} {Phys. Rev. B}\ }\textbf {\bibinfo {volume} {79}},\ \bibinfo
  {pages} {195438} (\bibinfo {year} {2009})}\BibitemShut {NoStop}%
\bibitem [{\citenamefont {Sanderson}\ \emph {et~al.}(2013)\citenamefont
  {Sanderson}, \citenamefont {Ang},\ and\ \citenamefont
  {Zhang}}]{sanderson2013Klein}%
  \BibitemOpen
  \bibfield  {author} {\bibinfo {author} {\bibfnamefont {M.}~\bibnamefont
  {Sanderson}}, \bibinfo {author} {\bibfnamefont {Y.~S.}\ \bibnamefont {Ang}},\
  and\ \bibinfo {author} {\bibfnamefont {C.}~\bibnamefont {Zhang}},\ }\bibfield
   {title} {\bibinfo {title} {Klein tunneling and cone transport in aa-stacked
  bilayer graphene},\ }\href {https://doi.org/10.1103/PhysRevB.88.245404}
  {\bibfield  {journal} {\bibinfo  {journal} {Phys. Rev. B}\ }\textbf {\bibinfo
  {volume} {88}},\ \bibinfo {pages} {245404} (\bibinfo {year}
  {2013})}\BibitemShut {NoStop}%
\bibitem [{\citenamefont {Greenaway}\ \emph {et~al.}(2015)\citenamefont
  {Greenaway}, \citenamefont {Vdovin}, \citenamefont {Mishchenko},
  \citenamefont {Makarovsky}, \citenamefont {Patan{\`e}}, \citenamefont
  {Wallbank}, \citenamefont {Cao}, \citenamefont {Kretinin}, \citenamefont
  {Zhu}, \citenamefont {Morozov} \emph {et~al.}}]{greenaway2015resonant}%
  \BibitemOpen
  \bibfield  {author} {\bibinfo {author} {\bibfnamefont {M.~T.}\ \bibnamefont
  {Greenaway}}, \bibinfo {author} {\bibfnamefont {E.~E.}\ \bibnamefont
  {Vdovin}}, \bibinfo {author} {\bibfnamefont {A.}~\bibnamefont {Mishchenko}},
  \bibinfo {author} {\bibfnamefont {O.}~\bibnamefont {Makarovsky}}, \bibinfo
  {author} {\bibfnamefont {A.}~\bibnamefont {Patan{\`e}}}, \bibinfo {author}
  {\bibfnamefont {J.}~\bibnamefont {Wallbank}}, \bibinfo {author}
  {\bibfnamefont {Y.}~\bibnamefont {Cao}}, \bibinfo {author} {\bibfnamefont
  {A.}~\bibnamefont {Kretinin}}, \bibinfo {author} {\bibfnamefont
  {M.}~\bibnamefont {Zhu}}, \bibinfo {author} {\bibfnamefont {S.}~\bibnamefont
  {Morozov}}, \emph {et~al.},\ }\bibfield  {title} {\bibinfo {title} {Resonant
  tunnelling between the chiral {L}andau states of twisted graphene lattices},\
  }\href {https://doi.org/https://doi.org/10.1038/nphys3507} {\bibfield
  {journal} {\bibinfo  {journal} {Nat. Phys.}\ }\textbf {\bibinfo {volume}
  {11}},\ \bibinfo {pages} {1057} (\bibinfo {year} {2015})}\BibitemShut
  {NoStop}%
\bibitem [{\citenamefont {Low}\ \emph {et~al.}(2009)\citenamefont {Low},
  \citenamefont {Hong}, \citenamefont {Appenzeller}, \citenamefont {Datta},\
  and\ \citenamefont {Lundstrom}}]{low2009conductance}%
  \BibitemOpen
  \bibfield  {author} {\bibinfo {author} {\bibfnamefont {T.}~\bibnamefont
  {Low}}, \bibinfo {author} {\bibfnamefont {S.}~\bibnamefont {Hong}}, \bibinfo
  {author} {\bibfnamefont {J.}~\bibnamefont {Appenzeller}}, \bibinfo {author}
  {\bibfnamefont {S.}~\bibnamefont {Datta}},\ and\ \bibinfo {author}
  {\bibfnamefont {M.~S.}\ \bibnamefont {Lundstrom}},\ }\bibfield  {title}
  {\bibinfo {title} {Conductance asymmetry of graphene pn junction},\ }\href
  {https://doi.org/10.1109/TED.2009.2017646} {\bibfield  {journal} {\bibinfo
  {journal} {IEEE Trans. Electron Devices}\ }\textbf {\bibinfo {volume} {56}},\
  \bibinfo {pages} {1292} (\bibinfo {year} {2009})}\BibitemShut {NoStop}%
\bibitem [{\citenamefont {Perconte}\ \emph {et~al.}(2018)\citenamefont
  {Perconte}, \citenamefont {Cuellar}, \citenamefont {Moreau-Luchaire},
  \citenamefont {Piquemal-Banci}, \citenamefont {Galceran}, \citenamefont
  {Kidambi}, \citenamefont {Martin}, \citenamefont {Hofmann}, \citenamefont
  {Bernard}, \citenamefont {Dlubak} \emph {et~al.}}]{perconte2018tunable}%
  \BibitemOpen
  \bibfield  {author} {\bibinfo {author} {\bibfnamefont {D.}~\bibnamefont
  {Perconte}}, \bibinfo {author} {\bibfnamefont {F.~A.}\ \bibnamefont
  {Cuellar}}, \bibinfo {author} {\bibfnamefont {C.}~\bibnamefont
  {Moreau-Luchaire}}, \bibinfo {author} {\bibfnamefont {M.}~\bibnamefont
  {Piquemal-Banci}}, \bibinfo {author} {\bibfnamefont {R.}~\bibnamefont
  {Galceran}}, \bibinfo {author} {\bibfnamefont {P.~R.}\ \bibnamefont
  {Kidambi}}, \bibinfo {author} {\bibfnamefont {M.-B.}\ \bibnamefont {Martin}},
  \bibinfo {author} {\bibfnamefont {S.}~\bibnamefont {Hofmann}}, \bibinfo
  {author} {\bibfnamefont {R.}~\bibnamefont {Bernard}}, \bibinfo {author}
  {\bibfnamefont {B.}~\bibnamefont {Dlubak}}, \emph {et~al.},\ }\bibfield
  {title} {\bibinfo {title} {Tunable {K}lein-like tunnelling of
  high-temperature superconducting pairs into graphene},\ }\href
  {https://doi.org/https://doi.org/10.1038/nphys4278} {\bibfield  {journal}
  {\bibinfo  {journal} {Nat. Phys.}\ }\textbf {\bibinfo {volume} {14}},\
  \bibinfo {pages} {25} (\bibinfo {year} {2018})}\BibitemShut {NoStop}%
\bibitem [{\citenamefont {Katsnelson}\ \emph {et~al.}(2006)\citenamefont
  {Katsnelson}, \citenamefont {Novoselov},\ and\ \citenamefont
  {Geim}}]{katsnelson2006chiral}%
  \BibitemOpen
  \bibfield  {author} {\bibinfo {author} {\bibfnamefont {M.~I.}\ \bibnamefont
  {Katsnelson}}, \bibinfo {author} {\bibfnamefont {K.~S.}\ \bibnamefont
  {Novoselov}},\ and\ \bibinfo {author} {\bibfnamefont {A.~K.}\ \bibnamefont
  {Geim}},\ }\bibfield  {title} {\bibinfo {title} {Chiral tunnelling and the
  {K}lein paradox in graphene},\ }\href
  {https://doi.org/https://doi.org/10.1038/nphys384} {\bibfield  {journal}
  {\bibinfo  {journal} {Nat. Phys.}\ }\textbf {\bibinfo {volume} {2}},\
  \bibinfo {pages} {620} (\bibinfo {year} {2006})}\BibitemShut {NoStop}%
\bibitem [{\citenamefont {Allain}\ and\ \citenamefont
  {Fuchs}(2011)}]{allain2011Klein}%
  \BibitemOpen
  \bibfield  {author} {\bibinfo {author} {\bibfnamefont {P.~E.}\ \bibnamefont
  {Allain}}\ and\ \bibinfo {author} {\bibfnamefont {J.~N.}\ \bibnamefont
  {Fuchs}},\ }\bibfield  {title} {\bibinfo {title} {Klein tunneling in
  graphene: optics with massless electrons},\ }\href
  {https://doi.org/https://doi.org/10.1140/epjb/e2011-20351-3} {\bibfield
  {journal} {\bibinfo  {journal} {Eur. Phys. J. B}\ }\textbf {\bibinfo {volume}
  {83}},\ \bibinfo {pages} {301} (\bibinfo {year} {2011})}\BibitemShut
  {NoStop}%
\bibitem [{\citenamefont {Beenakker}(2008)}]{RevModPhys.80.1337}%
  \BibitemOpen
  \bibfield  {author} {\bibinfo {author} {\bibfnamefont {C.~W.~J.}\
  \bibnamefont {Beenakker}},\ }\bibfield  {title} {\bibinfo {title}
  {Colloquium: {A}ndreev reflection and {K}lein tunneling in graphene},\ }\href
  {https://doi.org/10.1103/RevModPhys.80.1337} {\bibfield  {journal} {\bibinfo
  {journal} {Rev. Mod. Phys.}\ }\textbf {\bibinfo {volume} {80}},\ \bibinfo
  {pages} {1337} (\bibinfo {year} {2008})}\BibitemShut {NoStop}%
\bibitem [{\citenamefont {Avishai}\ and\ \citenamefont
  {Band}(2021)}]{avishai2021chiral}%
  \BibitemOpen
  \bibfield  {author} {\bibinfo {author} {\bibfnamefont {Y.}~\bibnamefont
  {Avishai}}\ and\ \bibinfo {author} {\bibfnamefont {Y.~B.}\ \bibnamefont
  {Band}},\ }\bibfield  {title} {\bibinfo {title} {Chiral tunneling in
  single-layer graphene with {R}ashba spin-orbit coupling: Spin currents},\
  }\href {https://doi.org/10.1103/PhysRevB.103.134445} {\bibfield  {journal}
  {\bibinfo  {journal} {Phys. Rev. B}\ }\textbf {\bibinfo {volume} {103}},\
  \bibinfo {pages} {134445} (\bibinfo {year} {2021})}\BibitemShut {NoStop}%
\bibitem [{\citenamefont {Avishai}\ and\ \citenamefont
  {Band}(2020)}]{avishai2020Klein}%
  \BibitemOpen
  \bibfield  {author} {\bibinfo {author} {\bibfnamefont {Y.}~\bibnamefont
  {Avishai}}\ and\ \bibinfo {author} {\bibfnamefont {Y.~B.}\ \bibnamefont
  {Band}},\ }\bibfield  {title} {\bibinfo {title} {Klein bound states in
  single-layer graphene},\ }\href {https://doi.org/10.1103/PhysRevB.102.085435}
  {\bibfield  {journal} {\bibinfo  {journal} {Phys. Rev. B}\ }\textbf {\bibinfo
  {volume} {102}},\ \bibinfo {pages} {085435} (\bibinfo {year}
  {2020})}\BibitemShut {NoStop}%
\bibitem [{\citenamefont {Rodrigues~da Silva}\ \emph
  {et~al.}(2019)\citenamefont {Rodrigues~da Silva}, \citenamefont {Barros},
  \citenamefont {Barbosa}, \citenamefont {Ramos} \emph {et~al.}}]{da2019Klein}%
  \BibitemOpen
  \bibfield  {author} {\bibinfo {author} {\bibfnamefont {A.~F.~M.}\
  \bibnamefont {Rodrigues~da Silva}}, \bibinfo {author} {\bibfnamefont
  {M.~S.~M.}\ \bibnamefont {Barros}}, \bibinfo {author} {\bibfnamefont
  {A.~L.~R.}\ \bibnamefont {Barbosa}}, \bibinfo {author} {\bibfnamefont {J.~G.
  G.~S.}\ \bibnamefont {Ramos}}, \emph {et~al.},\ }\bibfield  {title} {\bibinfo
  {title} {Klein paradox in chaotic {D}irac billiards},\ }\href
  {https://doi.org/https://doi.org/10.1016/j.aop.2019.03.011} {\bibfield
  {journal} {\bibinfo  {journal} {Ann. Phys.}\ }\textbf {\bibinfo {volume}
  {405}},\ \bibinfo {pages} {256} (\bibinfo {year} {2019})}\BibitemShut
  {NoStop}%
\bibitem [{\citenamefont {Szab{\'o}}\ \emph {et~al.}(2013)\citenamefont
  {Szab{\'o}}, \citenamefont {Benedict}, \citenamefont {Czirj{\'a}k},\ and\
  \citenamefont {F{\"o}ldi}}]{szabo2013relativistic}%
  \BibitemOpen
  \bibfield  {author} {\bibinfo {author} {\bibfnamefont {L.~Z.}\ \bibnamefont
  {Szab{\'o}}}, \bibinfo {author} {\bibfnamefont {M.~G.}\ \bibnamefont
  {Benedict}}, \bibinfo {author} {\bibfnamefont {A.}~\bibnamefont
  {Czirj{\'a}k}},\ and\ \bibinfo {author} {\bibfnamefont {P.}~\bibnamefont
  {F{\"o}ldi}},\ }\bibfield  {title} {\bibinfo {title} {Relativistic electron
  transport through an oscillating barrier: Wave-packet generation and
  {F}ano-type resonances},\ }\href {https://doi.org/10.1103/PhysRevB.88.075438}
  {\bibfield  {journal} {\bibinfo  {journal} {Phys. Rev. B}\ }\textbf {\bibinfo
  {volume} {88}},\ \bibinfo {pages} {075438} (\bibinfo {year}
  {2013})}\BibitemShut {NoStop}%
\bibitem [{\citenamefont {Prada}\ \emph {et~al.}(2010)\citenamefont {Prada},
  \citenamefont {San-Jose},\ and\ \citenamefont {Brey}}]{prada2010zero}%
  \BibitemOpen
  \bibfield  {author} {\bibinfo {author} {\bibfnamefont {E.}~\bibnamefont
  {Prada}}, \bibinfo {author} {\bibfnamefont {P.}~\bibnamefont {San-Jose}},\
  and\ \bibinfo {author} {\bibfnamefont {L.}~\bibnamefont {Brey}},\ }\bibfield
  {title} {\bibinfo {title} {Zero {L}andau level in folded graphene
  nanoribbons},\ }\href {https://doi.org/10.1103/PhysRevLett.105.106802}
  {\bibfield  {journal} {\bibinfo  {journal} {Phys. Rev. Lett.}\ }\textbf
  {\bibinfo {volume} {105}},\ \bibinfo {pages} {106802} (\bibinfo {year}
  {2010})}\BibitemShut {NoStop}%
\bibitem [{\citenamefont {Contreras-Astorga}\ \emph {et~al.}(2020)\citenamefont
  {Contreras-Astorga}, \citenamefont {Correa},\ and\ \citenamefont
  {Jakubsk{\'{y}}}}]{Contreras-Astorga2020}%
  \BibitemOpen
  \bibfield  {author} {\bibinfo {author} {\bibfnamefont {A.}~\bibnamefont
  {Contreras-Astorga}}, \bibinfo {author} {\bibfnamefont {F.}~\bibnamefont
  {Correa}},\ and\ \bibinfo {author} {\bibfnamefont {V.}~\bibnamefont
  {Jakubsk{\'{y}}}},\ }\bibfield  {title} {\bibinfo {title} {{Super-Klein
  tunneling of Dirac fermions through electrostatic gratings in graphene}},\
  }\href {https://doi.org/10.1103/PhysRevB.102.115429} {\bibfield  {journal}
  {\bibinfo  {journal} {Phys. Rev. B}\ }\textbf {\bibinfo {volume} {102}},\
  \bibinfo {pages} {115429} (\bibinfo {year} {2020})}\BibitemShut {NoStop}%
\bibitem [{\citenamefont {Lee}\ \emph {et~al.}(2019)\citenamefont {Lee},
  \citenamefont {Stanev}, \citenamefont {Zhang}, \citenamefont {Stasak},
  \citenamefont {Flowers}, \citenamefont {Higgins}, \citenamefont {Dai},
  \citenamefont {Blum}, \citenamefont {Pan}, \citenamefont {Yakovenko} \emph
  {et~al.}}]{lee2019perfect}%
  \BibitemOpen
  \bibfield  {author} {\bibinfo {author} {\bibfnamefont {S.}~\bibnamefont
  {Lee}}, \bibinfo {author} {\bibfnamefont {V.}~\bibnamefont {Stanev}},
  \bibinfo {author} {\bibfnamefont {X.}~\bibnamefont {Zhang}}, \bibinfo
  {author} {\bibfnamefont {D.}~\bibnamefont {Stasak}}, \bibinfo {author}
  {\bibfnamefont {J.}~\bibnamefont {Flowers}}, \bibinfo {author} {\bibfnamefont
  {J.~S.}\ \bibnamefont {Higgins}}, \bibinfo {author} {\bibfnamefont
  {S.}~\bibnamefont {Dai}}, \bibinfo {author} {\bibfnamefont {T.}~\bibnamefont
  {Blum}}, \bibinfo {author} {\bibfnamefont {X.}~\bibnamefont {Pan}}, \bibinfo
  {author} {\bibfnamefont {V.~M.}\ \bibnamefont {Yakovenko}}, \emph {et~al.},\
  }\bibfield  {title} {\bibinfo {title} {Perfect {A}ndreev reflection due to
  the {K}lein paradox in a topological superconducting state},\ }\href
  {https://doi.org/https://doi.org/10.1038/s41586-019-1305-1} {\bibfield
  {journal} {\bibinfo  {journal} {Nature}\ }\textbf {\bibinfo {volume} {570}},\
  \bibinfo {pages} {344} (\bibinfo {year} {2019})}\BibitemShut {NoStop}%
\bibitem [{\citenamefont {Kretinin}\ \emph {et~al.}(2013)\citenamefont
  {Kretinin}, \citenamefont {Yu}, \citenamefont {Jalil}, \citenamefont {Cao},
  \citenamefont {Withers}, \citenamefont {Mishchenko}, \citenamefont
  {Katsnelson}, \citenamefont {Novoselov}, \citenamefont {Geim},\ and\
  \citenamefont {Guinea}}]{PhysRevB.88.165427}%
  \BibitemOpen
  \bibfield  {author} {\bibinfo {author} {\bibfnamefont {A.}~\bibnamefont
  {Kretinin}}, \bibinfo {author} {\bibfnamefont {G.~L.}\ \bibnamefont {Yu}},
  \bibinfo {author} {\bibfnamefont {R.}~\bibnamefont {Jalil}}, \bibinfo
  {author} {\bibfnamefont {Y.}~\bibnamefont {Cao}}, \bibinfo {author}
  {\bibfnamefont {F.}~\bibnamefont {Withers}}, \bibinfo {author} {\bibfnamefont
  {A.}~\bibnamefont {Mishchenko}}, \bibinfo {author} {\bibfnamefont {M.~I.}\
  \bibnamefont {Katsnelson}}, \bibinfo {author} {\bibfnamefont {K.~S.}\
  \bibnamefont {Novoselov}}, \bibinfo {author} {\bibfnamefont {A.~K.}\
  \bibnamefont {Geim}},\ and\ \bibinfo {author} {\bibfnamefont
  {F.}~\bibnamefont {Guinea}},\ }\bibfield  {title} {\bibinfo {title} {Quantum
  capacitance measurements of electron-hole asymmetry and next-nearest-neighbor
  hopping in graphene},\ }\href {https://doi.org/10.1103/PhysRevB.88.165427}
  {\bibfield  {journal} {\bibinfo  {journal} {Phys. Rev. B}\ }\textbf {\bibinfo
  {volume} {88}},\ \bibinfo {pages} {165427} (\bibinfo {year}
  {2013})}\BibitemShut {NoStop}%
\bibitem [{\citenamefont {Zhang}\ and\ \citenamefont
  {Yang}(2018)}]{PhysRevB.97.035420}%
  \BibitemOpen
  \bibfield  {author} {\bibinfo {author} {\bibfnamefont {S.-H.}\ \bibnamefont
  {Zhang}}\ and\ \bibinfo {author} {\bibfnamefont {W.}~\bibnamefont {Yang}},\
  }\bibfield  {title} {\bibinfo {title} {Perfect transmission at oblique
  incidence by trigonal warping in graphene p-n junctions},\ }\href
  {https://doi.org/10.1103/PhysRevB.97.035420} {\bibfield  {journal} {\bibinfo
  {journal} {Phys. Rev. B}\ }\textbf {\bibinfo {volume} {97}},\ \bibinfo
  {pages} {035420} (\bibinfo {year} {2018})}\BibitemShut {NoStop}%
\bibitem [{\citenamefont {Pereira}\ \emph {et~al.}(2010)\citenamefont
  {Pereira}, \citenamefont {Peeters}, \citenamefont {Chaves},\ and\
  \citenamefont {Farias}}]{Pereira_2010}%
  \BibitemOpen
  \bibfield  {author} {\bibinfo {author} {\bibfnamefont {J.~M.}\ \bibnamefont
  {Pereira}}, \bibinfo {author} {\bibfnamefont {F.~M.}\ \bibnamefont
  {Peeters}}, \bibinfo {author} {\bibfnamefont {A.}~\bibnamefont {Chaves}},\
  and\ \bibinfo {author} {\bibfnamefont {G.~A.}\ \bibnamefont {Farias}},\
  }\bibfield  {title} {\bibinfo {title} {Klein tunneling in single and multiple
  barriers in graphene},\ }\href
  {https://doi.org/10.1088/0268-1242/25/3/033002} {\bibfield  {journal}
  {\bibinfo  {journal} {Semicond. Sci. Technol.}\ }\textbf {\bibinfo {volume}
  {25}},\ \bibinfo {pages} {033002} (\bibinfo {year} {2010})}\BibitemShut
  {NoStop}%
\bibitem [{\citenamefont {Logemann}\ \emph {et~al.}(2015)\citenamefont
  {Logemann}, \citenamefont {Reijnders}, \citenamefont {Tudorovskiy},
  \citenamefont {Katsnelson},\ and\ \citenamefont {Yuan}}]{PhysRevB.91.045420}%
  \BibitemOpen
  \bibfield  {author} {\bibinfo {author} {\bibfnamefont {R.}~\bibnamefont
  {Logemann}}, \bibinfo {author} {\bibfnamefont {K.~J.~A.}\ \bibnamefont
  {Reijnders}}, \bibinfo {author} {\bibfnamefont {T.}~\bibnamefont
  {Tudorovskiy}}, \bibinfo {author} {\bibfnamefont {M.~I.}\ \bibnamefont
  {Katsnelson}},\ and\ \bibinfo {author} {\bibfnamefont {S.}~\bibnamefont
  {Yuan}},\ }\bibfield  {title} {\bibinfo {title} {Modeling {K}lein tunneling
  and caustics of electron waves in graphene},\ }\href
  {https://doi.org/10.1103/PhysRevB.91.045420} {\bibfield  {journal} {\bibinfo
  {journal} {Phys. Rev. B}\ }\textbf {\bibinfo {volume} {91}},\ \bibinfo
  {pages} {045420} (\bibinfo {year} {2015})}\BibitemShut {NoStop}%
\bibitem [{\citenamefont {Pereira~Jr}\ \emph {et~al.}(2008)\citenamefont
  {Pereira~Jr}, \citenamefont {Peeters}, \citenamefont {Costa~Filho},\ and\
  \citenamefont {Farias}}]{Pereira_Jr_2008}%
  \BibitemOpen
  \bibfield  {author} {\bibinfo {author} {\bibfnamefont {J.~M.}\ \bibnamefont
  {Pereira~Jr}}, \bibinfo {author} {\bibfnamefont {F.~M.}\ \bibnamefont
  {Peeters}}, \bibinfo {author} {\bibfnamefont {R.~N.}\ \bibnamefont
  {Costa~Filho}},\ and\ \bibinfo {author} {\bibfnamefont {G.~A.}\ \bibnamefont
  {Farias}},\ }\bibfield  {title} {\bibinfo {title} {Valley polarization due to
  trigonal warping on tunneling electrons in graphene},\ }\href
  {https://doi.org/10.1088/0953-8984/21/4/045301} {\bibfield  {journal}
  {\bibinfo  {journal} {J. Phys.: Condens. Matter}\ }\textbf {\bibinfo {volume}
  {21}},\ \bibinfo {pages} {045301} (\bibinfo {year} {2008})}\BibitemShut
  {NoStop}%
\bibitem [{\citenamefont {Bahat-Treidel}\ \emph {et~al.}(2010)\citenamefont
  {Bahat-Treidel}, \citenamefont {Peleg}, \citenamefont {Grobman},
  \citenamefont {Shapira}, \citenamefont {Segev},\ and\ \citenamefont
  {Pereg-Barnea}}]{PhysRevLett.104.063901}%
  \BibitemOpen
  \bibfield  {author} {\bibinfo {author} {\bibfnamefont {O.}~\bibnamefont
  {Bahat-Treidel}}, \bibinfo {author} {\bibfnamefont {O.}~\bibnamefont
  {Peleg}}, \bibinfo {author} {\bibfnamefont {M.}~\bibnamefont {Grobman}},
  \bibinfo {author} {\bibfnamefont {N.}~\bibnamefont {Shapira}}, \bibinfo
  {author} {\bibfnamefont {M.}~\bibnamefont {Segev}},\ and\ \bibinfo {author}
  {\bibfnamefont {T.}~\bibnamefont {Pereg-Barnea}},\ }\bibfield  {title}
  {\bibinfo {title} {Klein tunneling in deformed honeycomb lattices},\ }\href
  {https://doi.org/10.1103/PhysRevLett.104.063901} {\bibfield  {journal}
  {\bibinfo  {journal} {Phys. Rev. Lett.}\ }\textbf {\bibinfo {volume} {104}},\
  \bibinfo {pages} {063901} (\bibinfo {year} {2010})}\BibitemShut {NoStop}%
\bibitem [{\citenamefont {Iorio}\ \emph {et~al.}(2018)\citenamefont {Iorio},
  \citenamefont {Pais}, \citenamefont {Elmashad}, \citenamefont {Ali},
  \citenamefont {Faizal},\ and\ \citenamefont
  {Abou-Salem}}]{doi:10.1142/S0218271818500803}%
  \BibitemOpen
  \bibfield  {author} {\bibinfo {author} {\bibfnamefont {A.}~\bibnamefont
  {Iorio}}, \bibinfo {author} {\bibfnamefont {P.}~\bibnamefont {Pais}},
  \bibinfo {author} {\bibfnamefont {I.~A.}\ \bibnamefont {Elmashad}}, \bibinfo
  {author} {\bibfnamefont {A.~F.}\ \bibnamefont {Ali}}, \bibinfo {author}
  {\bibfnamefont {M.}~\bibnamefont {Faizal}},\ and\ \bibinfo {author}
  {\bibfnamefont {L.~I.}\ \bibnamefont {Abou-Salem}},\ }\bibfield  {title}
  {\bibinfo {title} {Generalized {D}irac structure beyond the linear regime in
  graphene},\ }\href {https://doi.org/10.1142/S0218271818500803} {\bibfield
  {journal} {\bibinfo  {journal} {Int. J. Mod. Phys. D}\ }\textbf {\bibinfo
  {volume} {27}},\ \bibinfo {pages} {1850080} (\bibinfo {year}
  {2018})}\BibitemShut {NoStop}%
\bibitem [{\citenamefont {Amelino-Camelia}(2013)}]{amelino2013quantum}%
  \BibitemOpen
  \bibfield  {author} {\bibinfo {author} {\bibfnamefont {G.}~\bibnamefont
  {Amelino-Camelia}},\ }\bibfield  {title} {\bibinfo {title} {Quantum-spacetime
  phenomenology},\ }\href {https://doi.org/https://doi.org/10.12942/lrr-2013-5}
  {\bibfield  {journal} {\bibinfo  {journal} {Living Rev. Relativ.}\ }\textbf
  {\bibinfo {volume} {16}},\ \bibinfo {pages} {1} (\bibinfo {year}
  {2013})}\BibitemShut {NoStop}%
\bibitem [{\citenamefont {Hossenfelder}(2013)}]{hossenfelder2013minimal}%
  \BibitemOpen
  \bibfield  {author} {\bibinfo {author} {\bibfnamefont {S.}~\bibnamefont
  {Hossenfelder}},\ }\bibfield  {title} {\bibinfo {title} {Minimal length scale
  scenarios for quantum gravity},\ }\href
  {https://doi.org/https://doi.org/10.12942/lrr-2013-2} {\bibfield  {journal}
  {\bibinfo  {journal} {Living Rev. Relativ.}\ }\textbf {\bibinfo {volume}
  {16}},\ \bibinfo {pages} {1} (\bibinfo {year} {2013})}\BibitemShut {NoStop}%
\bibitem [{\citenamefont {Das}\ and\ \citenamefont
  {Vagenas}(2008)}]{PhysRevLett.101.221301}%
  \BibitemOpen
  \bibfield  {author} {\bibinfo {author} {\bibfnamefont {S.}~\bibnamefont
  {Das}}\ and\ \bibinfo {author} {\bibfnamefont {E.~C.}\ \bibnamefont
  {Vagenas}},\ }\bibfield  {title} {\bibinfo {title} {Universality of quantum
  gravity corrections},\ }\href
  {https://doi.org/10.1103/PhysRevLett.101.221301} {\bibfield  {journal}
  {\bibinfo  {journal} {Phys. Rev. Lett.}\ }\textbf {\bibinfo {volume} {101}},\
  \bibinfo {pages} {221301} (\bibinfo {year} {2008})}\BibitemShut {NoStop}%
\bibitem [{\citenamefont {Ali}\ \emph {et~al.}(2011)\citenamefont {Ali},
  \citenamefont {Das},\ and\ \citenamefont {Vagenas}}]{PhysRevD.84.044013}%
  \BibitemOpen
  \bibfield  {author} {\bibinfo {author} {\bibfnamefont {A.~F.}\ \bibnamefont
  {Ali}}, \bibinfo {author} {\bibfnamefont {S.}~\bibnamefont {Das}},\ and\
  \bibinfo {author} {\bibfnamefont {E.~C.}\ \bibnamefont {Vagenas}},\
  }\bibfield  {title} {\bibinfo {title} {Proposal for testing quantum gravity
  in the lab},\ }\href {https://doi.org/10.1103/PhysRevD.84.044013} {\bibfield
  {journal} {\bibinfo  {journal} {Phys. Rev. D}\ }\textbf {\bibinfo {volume}
  {84}},\ \bibinfo {pages} {044013} (\bibinfo {year} {2011})}\BibitemShut
  {NoStop}%
\bibitem [{\citenamefont {Goerbig}(2011)}]{RevModPhys.83.1193}%
  \BibitemOpen
  \bibfield  {author} {\bibinfo {author} {\bibfnamefont {M.~O.}\ \bibnamefont
  {Goerbig}},\ }\bibfield  {title} {\bibinfo {title} {Electronic properties of
  graphene in a strong magnetic field},\ }\href
  {https://doi.org/10.1103/RevModPhys.83.1193} {\bibfield  {journal} {\bibinfo
  {journal} {Rev. Mod. Phys.}\ }\textbf {\bibinfo {volume} {83}},\ \bibinfo
  {pages} {1193} (\bibinfo {year} {2011})}\BibitemShut {NoStop}%
\bibitem [{\citenamefont {Plochocka}\ \emph {et~al.}(2008)\citenamefont
  {Plochocka}, \citenamefont {Faugeras}, \citenamefont {Orlita}, \citenamefont
  {Sadowski}, \citenamefont {Martinez}, \citenamefont {Potemski}, \citenamefont
  {Goerbig}, \citenamefont {Fuchs}, \citenamefont {Berger},\ and\ \citenamefont
  {de~Heer}}]{PhysRevLett.100.087401}%
  \BibitemOpen
  \bibfield  {author} {\bibinfo {author} {\bibfnamefont {P.}~\bibnamefont
  {Plochocka}}, \bibinfo {author} {\bibfnamefont {C.}~\bibnamefont {Faugeras}},
  \bibinfo {author} {\bibfnamefont {M.}~\bibnamefont {Orlita}}, \bibinfo
  {author} {\bibfnamefont {M.~L.}\ \bibnamefont {Sadowski}}, \bibinfo {author}
  {\bibfnamefont {G.}~\bibnamefont {Martinez}}, \bibinfo {author}
  {\bibfnamefont {M.}~\bibnamefont {Potemski}}, \bibinfo {author}
  {\bibfnamefont {M.~O.}\ \bibnamefont {Goerbig}}, \bibinfo {author}
  {\bibfnamefont {J.~N.}\ \bibnamefont {Fuchs}}, \bibinfo {author}
  {\bibfnamefont {C.}~\bibnamefont {Berger}},\ and\ \bibinfo {author}
  {\bibfnamefont {W.~A.}\ \bibnamefont {de~Heer}},\ }\bibfield  {title}
  {\bibinfo {title} {High-energy limit of massless {D}irac fermions in
  multilayer graphene using magneto-optical transmission spectroscopy},\ }\href
  {https://doi.org/10.1103/PhysRevLett.100.087401} {\bibfield  {journal}
  {\bibinfo  {journal} {Phys. Rev. Lett.}\ }\textbf {\bibinfo {volume} {100}},\
  \bibinfo {pages} {087401} (\bibinfo {year} {2008})}\BibitemShut {NoStop}%
\bibitem [{\citenamefont {Amelino-Camelia}\ \emph {et~al.}(1997)\citenamefont
  {Amelino-Camelia}, \citenamefont {Ellis}, \citenamefont {Mavromatos},\ and\
  \citenamefont {Nanopoulos}}]{doi:10.1142/S0217751X97000566}%
  \BibitemOpen
  \bibfield  {author} {\bibinfo {author} {\bibfnamefont {G.}~\bibnamefont
  {Amelino-Camelia}}, \bibinfo {author} {\bibfnamefont {J.}~\bibnamefont
  {Ellis}}, \bibinfo {author} {\bibfnamefont {N.~E.}\ \bibnamefont
  {Mavromatos}},\ and\ \bibinfo {author} {\bibfnamefont {D.~V.}\ \bibnamefont
  {Nanopoulos}},\ }\bibfield  {title} {\bibinfo {title} {Distance measurement
  and wave dispersion in a {L}iouville-string approach to quantum gravity},\
  }\href {https://doi.org/10.1142/S0217751X97000566} {\bibfield  {journal}
  {\bibinfo  {journal} {Int. J. Mod. Phys. A}\ }\textbf {\bibinfo {volume}
  {12}},\ \bibinfo {pages} {607} (\bibinfo {year} {1997})}\BibitemShut
  {NoStop}%
\bibitem [{\citenamefont {Amelino-Camelia}(2004)}]{Amelino2004}%
  \BibitemOpen
  \bibfield  {author} {\bibinfo {author} {\bibfnamefont {G.}~\bibnamefont
  {Amelino-Camelia}},\ }\bibfield  {title} {\bibinfo {title} {Phenomenology of
  {P}lanck-scale {L}orentz-symmetry test theories},\ }\href
  {https://doi.org/10.1088/1367-2630/6/1/188} {\bibfield  {journal} {\bibinfo
  {journal} {New J. Phys.}\ }\textbf {\bibinfo {volume} {6}},\ \bibinfo {pages}
  {188} (\bibinfo {year} {2004})}\BibitemShut {NoStop}%
\bibitem [{\citenamefont {Aloisio}\ \emph {et~al.}(2000)\citenamefont
  {Aloisio}, \citenamefont {Blasi}, \citenamefont {Ghia},\ and\ \citenamefont
  {Grillo}}]{PhysRevD.62.053010}%
  \BibitemOpen
  \bibfield  {author} {\bibinfo {author} {\bibfnamefont {R.}~\bibnamefont
  {Aloisio}}, \bibinfo {author} {\bibfnamefont {P.}~\bibnamefont {Blasi}},
  \bibinfo {author} {\bibfnamefont {P.~L.}\ \bibnamefont {Ghia}},\ and\
  \bibinfo {author} {\bibfnamefont {A.~F.}\ \bibnamefont {Grillo}},\ }\bibfield
   {title} {\bibinfo {title} {Probing the structure of space-time with cosmic
  rays},\ }\href {https://doi.org/10.1103/PhysRevD.62.053010} {\bibfield
  {journal} {\bibinfo  {journal} {Phys. Rev. D}\ }\textbf {\bibinfo {volume}
  {62}},\ \bibinfo {pages} {053010} (\bibinfo {year} {2000})}\BibitemShut
  {NoStop}%
\bibitem [{\citenamefont {Das}\ \emph {et~al.}(2010)\citenamefont {Das},
  \citenamefont {Vagenas},\ and\ \citenamefont {Ali}}]{DAS2010407}%
  \BibitemOpen
  \bibfield  {author} {\bibinfo {author} {\bibfnamefont {S.}~\bibnamefont
  {Das}}, \bibinfo {author} {\bibfnamefont {E.~C.}\ \bibnamefont {Vagenas}},\
  and\ \bibinfo {author} {\bibfnamefont {A.~F.}\ \bibnamefont {Ali}},\
  }\bibfield  {title} {\bibinfo {title} {Discreteness of space from {GUP} ii:
  Relativistic wave equations},\ }\href
  {https://doi.org/https://doi.org/10.1016/j.physletb.2010.05.052} {\bibfield
  {journal} {\bibinfo  {journal} {Phys. Lett. B}\ }\textbf {\bibinfo {volume}
  {690}},\ \bibinfo {pages} {407} (\bibinfo {year} {2010})}\BibitemShut
  {NoStop}%
\bibitem [{\citenamefont {Ali}\ \emph {et~al.}(2009)\citenamefont {Ali},
  \citenamefont {Das},\ and\ \citenamefont {Vagenas}}]{ALI2009497}%
  \BibitemOpen
  \bibfield  {author} {\bibinfo {author} {\bibfnamefont {A.~F.}\ \bibnamefont
  {Ali}}, \bibinfo {author} {\bibfnamefont {S.}~\bibnamefont {Das}},\ and\
  \bibinfo {author} {\bibfnamefont {E.~C.}\ \bibnamefont {Vagenas}},\
  }\bibfield  {title} {\bibinfo {title} {Discreteness of space from the
  generalized uncertainty principle},\ }\href
  {https://doi.org/https://doi.org/10.1016/j.physletb.2009.06.061} {\bibfield
  {journal} {\bibinfo  {journal} {Phys. Lett. B}\ }\textbf {\bibinfo {volume}
  {678}},\ \bibinfo {pages} {497} (\bibinfo {year} {2009})}\BibitemShut
  {NoStop}%
\bibitem [{\citenamefont {Tusche}\ \emph {et~al.}(2015)\citenamefont {Tusche},
  \citenamefont {Krasyuk},\ and\ \citenamefont {Kirschner}}]{TUSCHE2015520}%
  \BibitemOpen
  \bibfield  {author} {\bibinfo {author} {\bibfnamefont {C.}~\bibnamefont
  {Tusche}}, \bibinfo {author} {\bibfnamefont {A.}~\bibnamefont {Krasyuk}},\
  and\ \bibinfo {author} {\bibfnamefont {J.}~\bibnamefont {Kirschner}},\
  }\bibfield  {title} {\bibinfo {title} {Spin resolved bandstructure imaging
  with a high resolution momentum microscope},\ }\href
  {https://doi.org/https://doi.org/10.1016/j.ultramic.2015.03.020} {\bibfield
  {journal} {\bibinfo  {journal} {Ultramicroscopy}\ }\textbf {\bibinfo {volume}
  {159}},\ \bibinfo {pages} {520} (\bibinfo {year} {2015})},\ \bibinfo {note}
  {special Issue: LEEM-PEEM 9}\BibitemShut {NoStop}%
\bibitem [{\citenamefont {Higuchi}\ \emph {et~al.}(2017)\citenamefont
  {Higuchi}, \citenamefont {Heide}, \citenamefont {Ullmann}, \citenamefont
  {Weber},\ and\ \citenamefont {Hommelhoff}}]{higuchi2017light}%
  \BibitemOpen
  \bibfield  {author} {\bibinfo {author} {\bibfnamefont {T.}~\bibnamefont
  {Higuchi}}, \bibinfo {author} {\bibfnamefont {C.}~\bibnamefont {Heide}},
  \bibinfo {author} {\bibfnamefont {K.}~\bibnamefont {Ullmann}}, \bibinfo
  {author} {\bibfnamefont {H.~B.}\ \bibnamefont {Weber}},\ and\ \bibinfo
  {author} {\bibfnamefont {P.}~\bibnamefont {Hommelhoff}},\ }\bibfield  {title}
  {\bibinfo {title} {Light-field-driven currents in graphene},\ }\href
  {https://doi.org/https://doi.org/10.1038/nature23900} {\bibfield  {journal}
  {\bibinfo  {journal} {Nature}\ }\textbf {\bibinfo {volume} {550}},\ \bibinfo
  {pages} {224} (\bibinfo {year} {2017})}\BibitemShut {NoStop}%
\bibitem [{\citenamefont {Hafez}\ \emph {et~al.}(2016)\citenamefont {Hafez},
  \citenamefont {Chai}, \citenamefont {Ibrahim}, \citenamefont {Mondal},
  \citenamefont {F{\'{e}}rachou}, \citenamefont {Ropagnol},\ and\ \citenamefont
  {Ozaki}}]{Hafez_2016}%
  \BibitemOpen
  \bibfield  {author} {\bibinfo {author} {\bibfnamefont {H.~A.}\ \bibnamefont
  {Hafez}}, \bibinfo {author} {\bibfnamefont {X.}~\bibnamefont {Chai}},
  \bibinfo {author} {\bibfnamefont {A.}~\bibnamefont {Ibrahim}}, \bibinfo
  {author} {\bibfnamefont {S.}~\bibnamefont {Mondal}}, \bibinfo {author}
  {\bibfnamefont {D.}~\bibnamefont {F{\'{e}}rachou}}, \bibinfo {author}
  {\bibfnamefont {X.}~\bibnamefont {Ropagnol}},\ and\ \bibinfo {author}
  {\bibfnamefont {T.}~\bibnamefont {Ozaki}},\ }\bibfield  {title} {\bibinfo
  {title} {Intense terahertz radiation and their applications},\ }\href
  {https://doi.org/10.1088/2040-8978/18/9/093004} {\bibfield  {journal}
  {\bibinfo  {journal} {J. Opt.}\ }\textbf {\bibinfo {volume} {18}},\ \bibinfo
  {pages} {093004} (\bibinfo {year} {2016})}\BibitemShut {NoStop}%
\bibitem [{\citenamefont {Cohen}\ and\ \citenamefont
  {Glashow}(2006)}]{PhysRevLett.97.021601}%
  \BibitemOpen
  \bibfield  {author} {\bibinfo {author} {\bibfnamefont {A.~G.}\ \bibnamefont
  {Cohen}}\ and\ \bibinfo {author} {\bibfnamefont {S.~L.}\ \bibnamefont
  {Glashow}},\ }\bibfield  {title} {\bibinfo {title} {Very special
  relativity},\ }\href {https://doi.org/10.1103/PhysRevLett.97.021601}
  {\bibfield  {journal} {\bibinfo  {journal} {Phys. Rev. Lett.}\ }\textbf
  {\bibinfo {volume} {97}},\ \bibinfo {pages} {021601} (\bibinfo {year}
  {2006})}\BibitemShut {NoStop}%
\bibitem [{\citenamefont {Shah}\ and\ \citenamefont
  {Ahsan}(2021)}]{doi:10.1142/S0219887821500481}%
  \BibitemOpen
  \bibfield  {author} {\bibinfo {author} {\bibfnamefont {N.~A.}\ \bibnamefont
  {Shah}}\ and\ \bibinfo {author} {\bibfnamefont {M.~A.~H.}\ \bibnamefont
  {Ahsan}},\ }\bibfield  {title} {\bibinfo {title} {Quantization of
  non-{A}belian gauge theory in graphene},\ }\href
  {https://doi.org/10.1142/S0219887821500481} {\bibfield  {journal} {\bibinfo
  {journal} {Int. J. Geom. Methods Mod. Phys.}\ }\textbf {\bibinfo {volume}
  {18}},\ \bibinfo {pages} {2150048} (\bibinfo {year} {2021})}\BibitemShut
  {NoStop}%
\bibitem [{\citenamefont {Mann}\ \emph {et~al.}(2021)\citenamefont {Mann},
  \citenamefont {Husin}, \citenamefont {Patel}, \citenamefont {Faizal},
  \citenamefont {Sulaksono},\ and\ \citenamefont {Suroso}}]{mann2021testing}%
  \BibitemOpen
  \bibfield  {author} {\bibinfo {author} {\bibfnamefont {R.~B.}\ \bibnamefont
  {Mann}}, \bibinfo {author} {\bibfnamefont {I.}~\bibnamefont {Husin}},
  \bibinfo {author} {\bibfnamefont {H.}~\bibnamefont {Patel}}, \bibinfo
  {author} {\bibfnamefont {M.}~\bibnamefont {Faizal}}, \bibinfo {author}
  {\bibfnamefont {A.}~\bibnamefont {Sulaksono}},\ and\ \bibinfo {author}
  {\bibfnamefont {A.}~\bibnamefont {Suroso}},\ }\bibfield  {title} {\bibinfo
  {title} {Testing short distance anisotropy in space},\ }\href
  {https://doi.org/https://doi.org/10.1038/s41598-021-86355-3} {\bibfield
  {journal} {\bibinfo  {journal} {Sci. Rep.}\ }\textbf {\bibinfo {volume}
  {11}},\ \bibinfo {pages} {1} (\bibinfo {year} {2021})}\BibitemShut {NoStop}%
\end{thebibliography}%
\end{document}